\documentclass[12pt,preprint]{aastex}

\shorttitle{UCD Radial Velocity Survey}
\shortauthors{Blake et al.}

\begin{document}


\title{The NIRSPEC Ultracool Dwarf Radial Velocity Survey}

\author{Cullen H. Blake}
\affil{Princeton University Department of Astrophysical Sciences, Peyton Hall, Ivy Lane, Princeton, NJ 08544}
\affil{Harvard-Smithsonian Center for Astrophysics, 60 Garden Street, Cambridge, MA, 02138}

\author{David Charbonneau}
\affil{Harvard-Smithsonian Center for Astrophysics, 60 Garden Street, Cambridge, MA, 02138}

\author{Russel J. White}
\affil{Georgia State University, Department of Physics and Astronomy, Atlanta, GA 30303}

\begin{abstract}

We report the results of an infrared Doppler survey designed to detect brown dwarf and giant planetary companions to a magnitude-limited sample of ultracool dwarfs. 
Using the NIRSPEC spectrograph on the Keck II telescope, we obtained approximately 600 radial velocity measurements over a period of six years for a sample of 59 late-M and L dwarfs spanning spectral types M8/L0 to L6. A subsample of 46 of our targets have been observed on three or more epochs. We rely on telluric CH$_{4}$ absorption features in the Earth's atmosphere as a simultaneous wavelength reference and exploit the rich set of CO 
absorption features found in the K-band spectra of cool stars and brown dwarfs to measure radial velocities and projected rotational velocities. For a bright, slowly rotating M dwarf standard we demonstrate a radial velocity precision of 50 m s$^{-1}$, and for slowly rotating L dwarfs we achieve a typical radial velocity precision of approximately 200 m s$^{-1}$. This precision is sufficient for the detection of close-in giant planetary companions to mid-L dwarfs as well as more equal mass spectroscopic binary systems with small separations ($a<2$ AU).  We present an orbital solution for the subdwarf binary LSR1610$-$0040 as well as an improved solution for the M/T binary 2M0320$-$04. We compare the distribution of our observed values for the projected rotational velocities, $V\sin{i}$, to those in the literature and find that our sample contains examples of slowly rotating mid-L dwarfs, which have not been seen in other surveys. We also combine our radial velocity measurements with distance estimates and proper motions from the literature to estimate the dispersion of the space velocities of the objects in our sample. Using a kinematic age estimate we conclude that our UCDs have an age of $5.0^{+0.7}_{-0.6}$ Gyr, similar to that of nearby sun-like stars. We simulate the efficiency with which we detect spectroscopic binaries and find that the rate of tight ($a<1$ AU) binaries in our sample is $2.5^{+8.6}_{-1.6}\%$, consistent with recent estimates in the literature of a tight binary fraction of $3-4\%$.

\end{abstract}


\section{Introduction}

Since the discovery of the prototypical ``hot Jupiter'' orbiting the
star 51 Pegasi in 1995, more than 450 extrasolar planets have been
identified. This flurry of discovery has been driven largely by
technological innovation and the development of new observational
techniques. Precise Doppler measurements have played a
particularly important role in the discovery of extrasolar
planets. The search for unseen orbiting companions to stars by
measuring the subtle Doppler shifts of stellar spectral lines is more
than a century old (e.g. \citealt{vogel1901}), and since the late 1800s the precision of these 
measurements has improved by more than four orders of magnitude. Since
the amplitude of the Doppler signal induced by an unseen companion is 
directly proportional to the companion's mass, the discovery
of extrasolar planets is a direct result of steady improvements to
measurement techniques that have been used for over a century to study
binary star systems.

Approximately $4\%$ of the known extrasolar planetary systems have low-mass star hosts (M0-M4; www.exoplanet.eu). The bias in Doppler surveys toward stars
more massive than M dwarfs is due largely to the technical limitations
of obtaining precise measurements of cool, intrinsically faint objects.  Understanding the rate of occurrence of planets orbiting these lowest mass stars and brown dwarfs,
collectively referred to as Ultracool Dwarfs (UCDs, spectral types later than M5), may have important
implications for theories of planet formation since the core accretion and disk instability formation scenarios make different predictions about the occurrence of planetary companions as a function of host mass \citep{boss2006}. While it
has been shown that early M dwarfs have a relative paucity of close in 
Jupiter-mass planets \citep{endl2006,johnson2007}, we are just now beginning to understand the 
occurrence of super-Earth ($M_p<10 M_{\earth}$) planets to mid-M stars or any type of
planet orbiting late-M or L dwarfs. Doppler planet surveys generally include no UCD targets \citep{bailey2009}, so little is known about the rate at which planetary companions accompany the lowest mass stars and brown dwarfs. There are some initial
indications from microlensing surveys that sub-Jupiter-mass companions
orbiting at several AU from M dwarfs may be common \citep{gould2006a}, but these findings are not yet statistically robust. Several examples of possible planetary companions to brown dwarfs do exist, including a $16-20M_{\rm{J}}$ companion found orbiting a young brown dwarf in a Doppler survey conducted by \citet{joergens2007}, a giant planetary companion to 2M1207$-$39  found by \citet{chauvin2005} in a direct imaging survey, and a possible super-Earth orbiting the UCD MOA-2007-BLG-192 detected via microlensing \citep{bennett2008}.

There is strong observational evidence for the initial stages of planet formation around young UCDs in the form of a high disk fraction \citep{luhman2005} and the formation of silicate grains \citep{apai2005}. The formation of planets around UCDs with $M_{*}<0.1M_{\sun}$ has been modeled by \citet{payne2007} who found that, depending on the mass of the protoplanetary disk, super-Earths up to $5M_{\earth}$ may be relatively common ($10\%$ of UCDs) if disk mass scales linearly with host mass. The interaction of a planet with a gaseous disk leads to gravitational torques that can cause a planet to lose angular momentum and migrate within the disk. For small planets ($M_{p}<<M_{\rm{J}}$), the disk-planet interactions are linear, resulting in a Type I migration that can cause  very rapid inward movement. More massive planets may be able to open a gap in the protoplanetary disk and then torques between the planet and the inner and outer edges of the gap may cause the planet to slowly migrate inward through Type II migration. UCD planets may be found at relatively large separations ($a>1$~AU) since inward migration to short orbital periods through Type II migration is not expected to be efficient in UCD protoplanetary disks. The detection of a significant population of close-in companions to UCDs would place interesting constraints on planetary migration via the Type I mechanism since the rapid inward movements due to this mechanism are expected to cause protoplanets to fall into the star on short timescales. \citet{payne2007} predict that the formation of giant planetary companions to UCDs should be completely inhibited and propose that systems like 2M1207$-$39 form in a manner similar to that of binary stars.

Doppler planet searches are very sensitive to spectroscopic binary systems. Binary
star systems afford one of the few opportunities to directly measure
the masses and radii of stars and measurements of stars in binary
systems are an important component of the observational basis of our theoretical models of stellar structure and evolution. The are four different types of spectroscopic binary systems: single-lined (SB1), double-lined (SB2), single-lined eclipsing (SEB), and double-lined eclipsing (DEB). In an SB1 binary the reflex motion of the primary (more luminous and usually more massive)
star is observed and when combined with the orbital period and eccentricity can be
used to define a mass function, a transcendental equation involving
the masses of both components and the inclination. The inclination of the system is not known, so additional observations, such as measurements of the astrometric orbit, are required to determine the actual masses of both components. In an SB2 system the spectral lines of both the primary and secondary are observed and the ratio of the radial velocity semi-amplitudes of the components directly determines the ratio of their masses ($q=M_{2}/M_{1}=K_{1}/K_{2}$). Using a technique like TODCOR \citep{mazeh1994} to analyze observations of an SB2 system it may also be possible to simultaneously determine mass and flux ratios of the components of the system,  enabling direct tests of theoretical models of coeval
low-mass stars and brown dwarfs. As with an SB1 system, the system inclination is not known in an SB2, so individual masses can not be directly measured. In an SEB system light from the secondary is not detected, but its presence is inferred from both the reflex motion of the primary and the diminution in brightness that occurs as the secondary eclipses the primary. In these systems, of which the transiting extrasolar planets are a specific case, the orbital inclination is constrained by the fact that an eclipse occurs. The DEB is a rare, but very important, type of binary that allows for precise measurements of the masses and radii of both components of the binary. 

Very few UCD spectroscopic
binaries are known, meaning that theoretical models of low-mass stars
and brown dwarfs are relatively untested compared to models of
sun-like stars. In fact, there are significant ($5-10\%$) discrepancies between
models and mass and radius measurements of low-mass stars in eclipsing binary systems \citep{chabrier2007}. For UCDs, only one DEB system is known \citep{stassun2006}, serving as the lone observational benchmark for models of young brown dwarfs. In this young ($t<0.003$~Gyr) system, the measured radii are consistent with theoretical models but the estimated temperatures indicate that the less massive component is actually hotter, contrary to theoretical predictions for coeval brown dwarfs. Since UCDs that appear brightest in the sky are necessarily
close to the sun, direct imaging surveys have been very successful in
detecting UCD binaries with orbital separations $a>1$~AU. The orbital motions in these long-period visual binary systems can be observed (e.g. \citealt{bouy2004}, \citealt{dupuy2009}, \citealt{martinache2009}, \citealt{konopacky2010}), providing another opportunity to measure masses of UCDs. Discovering additional UCD binaries, particularly SB2 and DEB systems, is crucial for improving our understanding of stellar astrophysics at and below the bottom of the main sequence.

Observations of a large sample of binary systems also
enable tests of models of the formation history of UCDs. The process through
which UCDs form is not well understood, and the statistical properties
of the orbital separations and mass ratios of binary systems help
to test potential formation scenarios. For example, some models of UCD formation suggest that these objects undergo ejection from star formation regions before they have the opportunity to accrete enough mass to become main-sequence stars (\citealt{whitworth2006}, \citealt{luhman2006}). If this is the case, wide binaries ($a>15$~AU) with low binding energies are expected to be rare. Thanks to direct imaging surveys, the
binary fraction at large separations has been well studied. Compared to Sun-like stars, UCD binaries tend to have larger mass ratios ($q\sim1$) and smaller separations ($a\sim7$~AU) and very few wide ($a>15$~AU) systems are seen.
Most systems with small ($a<1$~AU) separations are not resolved by imaging surveys, but their relatively short periods (years instead of
decades) make them prime targets for Doppler surveys. Since no comprehensive Doppler
survey of UCDs has been carried out, our understanding of the overall
distribution of system properties is incomplete at small separations
\citep{burgasser2007b, allen2007}.

Obtaining precise Doppler measurements poses a significant technical challenge, particularly if the target is a UCD. In the 1970s astronomers realized that having a simultaneous wavelength calibrator that superimposes absorption features of known wavelengths onto a source spectrum provides a major advantage in terms of Doppler precision for slit spectrographs \citep{griffin1973}. Prior to the advent of the gas cell technique (HF by \citet{campbell1983} and later I$_{2}$), atomic and molecular absorption features in the Earth's atmosphere were used for this purpose. While not inherently stable like the gas in an absorption cell, telluric lines have been used to produce Doppler measurements with a month-to-month precision of 5-20 m s$^{-1}$ \citep{balthasar1982,smith1982,caccin1985,cochran1988,hatzes1993,figueira2010a}. After making corrections based on a simple model of atmospheric winds, \citet{figueira2010b} have demonstrated that atmospheric O$_2$ lines at optical wavelengths can be used to make RV measurements with a precision of 2 m s$^{-1}$ over timescales of years. Doppler measurements of Sun-like stars with a precision exceeding 1 m
s$^{-1}$ have been demonstrated using two different techniques for
calibrating high-resolution spectroscopic data. For
bright early-M dwarfs both the Thorium-Argon
(ThAr) emission lamps and simultaneous iodine (I$_{2}$) absorption cells
have been used to obtain RV precision of 3-5 m s$^{-1}$ \citep{endl2006, johnson2007, udry2007,zechmeister2009}. Due to their cool temperatures
($1200<T_{\rm{eff}}<2600$K) and small sizes ($R_{*}=0.1~R_{\sun}$), UCDs are intrinsically very faint at the
wavelengths where these measurements are made (400 to 700 nm), limiting
observations to only the few brightest targets on the sky. Making
precise Doppler measurements at near infrared (NIR) wavelengths is a very
attractive option for exploring the population of planets orbiting
UCDs.

 Until quite recently, the precision of NIR Doppler measurements of
UCDs has tended to lag behind that of RV measurements of Sun-like
stars by two orders of magnitude. This is due in part to the relative complexity and expense of NIR
echelle spectrographs, the relative faintness of cool dwarfs, and to the worse noise properties of
NIR detectors compared to CCDs. In addition, there has been a lack of suitable
wavelength references in the NIR. The I$_{2}$
cell is not effective at these wavelengths, but the Th lines in ThAr emission lamps may prove useful in $J$ and $H$ bands.
Both \citet{mahadevan2009} and \citet{reiners2010} provide a summary of future prospects for calibrating 
NIR echelle spectra using gas absorption cells and emission line lamps.  The NIR is replete with telluric absorption lines due to H$_{2}$O and CH$_{4}$. In this spectral region UCDs have a rich set of molecular absorption features that can potentially be used to make precise Doppler measurements. We have developed a technique that relies on telluric CH$_{4}$ absorption features as a simultaneous wavelength reference and exploits the rich set of CO absorption features found in the spectra of UCDs near 2.3$\mu$m to make Doppler measurements with a limiting precision of approximately 50 m s$^{-1}$. High-resolution NIR spectrographs that could make use of this technique are expected to be an important component of the suites of instruments on future large telescopes and \citet{ramsey2008} and \citet{erskine2005} report the development of new high-resolution instruments that should be able to achieve 10 m s$^{-1}$ precision at NIR wavelengths. Doppler measurements with a precision of $\sigma_{RV}<300$ m s$^{-1}$ have been demonstrated by \citet{martin2006}, \citet{blake2007a}, \citet{blake2008b}, \citet{prato2008},and \citet{zapatero2009}, and a Doppler precision in the range 5-20 m s$^{-1}$ has been demonstrated on short timescales using the CRIRES instrument on VLT by \citet{huelamo2008}, \citet{seifahrt2008}, and \citet{figueira2010a}. Recently, \citet{bean2009} used an NH$_3$ absorption cell with CRIRES to obtain RV measurements of bright ($K<8$) mid-M dwarfs in K band with a precision approaching 5 m s$^{-1}$. We do not expect to reach this impressive level of precision for two reasons: First, the resolution of NIRSPEC is R=25000, which is significantly worse than that of CRIRES (R=100000). Second, all but eight of the 59 targets in our survey have K magnitudes between 11 and 12.5, significantly fainter than the faintest targets reported in \citet{bean2009} or \citet{figueira2010a}.

We report the results of a Doppler survey of 59 UCDs using the NIRSPEC instrument on the Keck II telescope. Our observations span a period of six years and we demonstrate sensitivity to giant planetary companions as well as UCD-UCD binaries with small orbital separations. In Section 2 we describe our UCD sample and our NIRSPEC observations. In Section 3 we describe the details of our reduction and calibration of the NIRSPEC data. In Section 4 we describe our NIR Doppler technique, the expected precision, and potential sources of noise that limit the overall precision we achieve. In Section 5 we discuss the overall statistical properties of our Doppler measurements. In Section 6 we describe four individual RV variables and present orbital solutions. In Section 7 we describe the rotational and kinematic properties of our sample and compare the distributions of these values to those in the literature. In Section 8 we estimate the rate of tight ($a<1$AU) UCD binaries and simulate the sensitivity of our survey to giant planet companions.

\section{Sample Selection and Observations}

Thanks to all-sky NIR surveys such as 2MASS, SDSS, and DENIS, the L spectral class is a well-studied group of several hundred old low-mass stars and younger brown dwarfs. There is an inherent degeneracy between age and spectral type for these objects, but at field ages ($t>1$~Gyr) the
mid-L dwarfs are expected to be brown dwarfs, objects with masses below the minimum required for main sequence hydrogen burning, while late M and early L dwarfs may be very small hydrogen burning stars \citep{burrows2001}. The brown dwarfs slowly cool, radiating away their initial thermal energy over billions of years. With their cool temperatures, the atmospheres of UCDs contain a wide array of molecules, including TiO, VO, and CO, as well as dust particles, leading to complex NIR spectra \citep{rayner2009}.

We selected a sample of field UCD dwarfs brighter
than $K=13.0$ observable from Mauna Kea (DEC$>-40^{\circ}$). Our targets and their observed properties are listed in Table 1. The majority of our targets are classified as L dwarfs, though three may be classified as early-L or late-M depending on the spectral diagnostics used. Today, our sample contains more than $70\%$ 
of the known L dwarfs that satisfy our magnitude and declination limits (www.dwarfarchives.org), though a number of new L dwarfs were discovered during the course of our survey. Between March, 2003 and May,
2009 we collected approximately 600 individual observations of a sample of
59 UCDs using the NIRSPEC \citep{mclean1998} instrument on the
Keck II telescope. NIRSPEC is a high-resolution, cross-dispersed NIR
echelle spectrograph and is a powerful instrument for
high-resolution spectroscopy of cool stars and brown dwarfs. While NIRSPEC can be operated in conjunction with the AO system, our observations were obtained without AO since many of our targets are too faint for AO observations.
We selected the 3-pixel (0.432$\arcsec$) slit, the N7 order blocking
filter, the thin IR blocker, and a spectrograph configuration designed to place our desired
spectral region around the CO bandhead (2.285 to 2.318~$\micron$)
near the center of echelle order 33. This setup provided a resolution of
R=25000 and the 3-pixel slit was a good match for the typical
seeing in K-band at Mauna Kea. Over the course of our survey we
utilized this same spectrograph set-up and by using emission line
lamps we were able to adjust the echelle and cross disperser
angles in order to reproduce the positions of the echelle
orders to within $\pm0.07$~nm. NIRSPEC employs a 1024x1024 pixel ALADDIN InSb array with 27$\micron$ pixels and all of our science data were gathered using Fowler sampling (MCDS-16) readouts in order to reduce the read noise to the nominal level of 25~e$^{-}$.

We gathered observations of our targets in nod pairs where the target was nodded along the slit by
approximately 6$\arcsec$ between the first and second exposures of a
pair. This observing strategy facilitates the removal of sky emission
lines through the subtraction of consecutive 2D images. We selected integration
times so as to achieve a S/N per pixel of between 50 and 100 in each
of our individual extracted 1D spectra with exposure times ranging from 500 to 1200 seconds per nod position. On each night we gathered an extensive set of calibration data 
including a large number of flat field images and observations of bright, rapidly rotating A stars at a
range of airmasses. The A star spectra are free from stellar absorption features in our spectral region
and are therefore useful for monitoring changes in telluric absorption. During the course of our survey we obtained between two and 16 epochs of observations for each target in our sample. 
This inhomogeneous pattern of visits was determined in part by the scheduling of our observing time, but objects were also prioritized based on their projected rotational velocity, $V\sin{i}$. Typically, we observed objects with large $V\sin{i}$, which limits our Doppler precision, only twice, while objects exhibiting clear evidence for Doppler variations were observed at every opportunity.

\section{Data Reduction}

We reduced the NIRSPEC data and extracted spectra from order 33 using a set of custom IDL procedures developed for this survey to flat field the 2D spectra, trace the spectral orders, and extract 1D spectra.
For each observing session, which we defined as the period between physical movements of the internal 
components of the spectrograph, sets of 20 flat field images were median combined to produce a ``superflat''. As a result of computer or hardware problems, there were occasionally multiple observing sessions defined within a single observing night.
The individual flat fields have an integration time of 4s, resulting in an average signal of $4500$ ADU per pixel in order 33. The ALADDIN detector has a small dark current (0.2 e$^{-1}$ pixel$^{-1}$ s$^{-1}$) so we also gathered 4s dark frames for use in the creation of the superflats. Prior to median combination the individual flat fields were each normalized so as to compensate for changes in the overall flux levels due to warming of the flat field lamp. We used the superflat to trace echelle order 33 and to define the position of the order across the detector. Since we replicated the same spectrograph configuration during each observing session, the positions of the echelle orders are known to within a few pixels \textit{a priori} in all of our data.

We began the reduction and extraction procedures by subtracting nod pairs. Provided that changes in the detector or spectrograph properties were negligible over the timescale of the nod pair, this effectively removes the dark current and the bias level of the detector. If the sky brightness is not rapidly changing then night sky emission lines are also removed by pair subtraction. Following subtraction, the two resulting 2D difference images (A-B and B-A) were flat fielded using the superflat to compensate for sensitivity variations both across the order and at the pixel-to-pixel level, resulting in intensity rectified difference images. We trimmed 24 noisy columns from one edge of the detector, leaving a total of 1,000 columns. In our description of the data the echelle orders run roughly parallel to the rows of the detector. We extracted spectra from the intensity rectified difference images following the procedures outlined in \citet{horne1986}. The position of the spectrum across order 33 was determined by fitting a Gaussian in the spatial direction at each column and then fitting the resulting centers to a fourth-order polynomial with outlier rejection. Using the difference image, we built an empirical model of the spectral profile within order 33 in the spatial direction. This model 
accommodates smooth variations in the width or shape of the profile across the order and is normalized so that the integral of the spectral profile at each column is unity. At each column we fit this model spectral profile to the data by solving for the scale factor and an additive offset, to account for incomplete sky subtraction, that best fits the data in a least squares sense. The variance for each pixel was determined from the quoted gain and read noise estimated from a region of the NIRSPEC detector between the spectral orders. We found that the read noise estimated in this way was often close to 75 e$^{-}$, much larger than the quoted value of 25 e$^{-}$. The optimal estimate of the total flux at each of the 1,000 columns was determined by the scale factor of the best-fit model spectral profile. Each profile fit was conducted iteratively to  mitigate the effects of cosmic rays or bad pixels by rejecting outliers and then re-fitting. 

Some of our NIRSPEC data exhibit a significant additional noise. A transient pattern is sometimes seen in a single
quadrant of the NIRSPEC detector such that every eighth row has significantly enhanced noise. The phase of this pattern changes in time both within a night and between nights while the eight pixel periodicity remains fixed. These noisy rows run roughly parallel to the echelle orders so, depending on the phase of the pattern, they can have a significant impact on the S/N of the extracted data. The enhanced noise was seen in approximately 22$\%$ of our observations, though not always in the immediate vicinity of order 33. In a smaller subset of our data more complex noise patterns were seen with multiple patterns each having an eight pixel period. We visually inspected all of the individual extracted spectra and culled approximately $10\%$ as having poor S/N or severe noise problems due to order 33 falling along a particularly noisy row.

\section{Spectral Modeling}

We forward modeled the extracted spectra to measure the stellar radial velocity (RV) and projected rotational velocity ($V\sin{i}$). This procedure followed the methods described in \citet{blake2007a, blake2008b} and is similar to that used by \citet{butler1996} to obtain 3 m s$^{-1}$ Doppler precision at optical wavelengths using an I$_{_2}$ absorption cell. Our simultaneous calibrator is CH$_{4}$ located not in a cell but in Earth's atmosphere. The basis for our modeling procedure is the interpolation and convolution of high-resolution spectral models to fit the lower resolution NIRSPEC data, which we denote $D$, by minimizing $\chi^{2}$. The basic form of our model can be expressed

\begin{equation}
 M(\lambda) = \left(\left[L\left(\lambda \times \left(1+{v\over{c}}\right)\right)\star  K\right] \times T(\lambda)\right) \star LSF
\end{equation}
where $\star$ indicates convolution, $L(\lambda)$ is a high resolution UCD template, $K$ is the rotational broadening kernel, $T(\lambda)$ is the telluric spectrum, and $LSF$ is the spectrograph line spread function. We began with a library of high-resolution synthetic template spectra computed as described in \citet{marley2002} and \citet{saumon2008}. The models apply the condensation cloud model of \citet{ackerman2001} with a  sedimentation parameter of $f_{\rm sed}=3$, corresponding to a moderate amount of  condensate settling.  The models used here have solar metallicity \citep{lodders2003}, use the opacities described in \citet{freedman2008}, a fixed gravity of $\log g=5$ (cgs), and cover a range of $T_{\rm eff}$ from 1200 to  2400$\,$K.  The synthetic spectra provide monochromatic fluxes spaced $4.2 \times  10^{-6}\,\mu$m apart. We incorporated line broadening due to stellar rotation by convolving with the kernel defined by \citet{gray1992} using a linear limb-darkening parameter of 0.6 as appropriate for cool stars at infrared wavelengths \citep{claret2000}. We also used a high-resolution ($5\times 10^{-6} \mu$m~spacing) telluric spectrum derived from observations of the Sun provided by \citet{livingston1991}. Using quadratic interpolation we placed the synthetic UCD template and telluric model onto an evenly spaced wavelength grid with $5 \times  10^{-6}\mu$m spacing, about five times finer than the NIRSPEC data. We convolved the product of the rotationally-broadened UCD template and the telluric model with an estimate of the spectrograph Line Spread Function (LSF). Finally, we used quadratic interpolation to place the model on the lower resolution NIRSPEC wavelength grid, which we define through polynomial mapping of pixel to wavelength. In total the model of each individual spectrum has the following 11 free parameters: four for the polynomial mapping of pixel to wavelength, one for an overall flux scaling, four for a flux gradient across the spectrum, one for the LSF FWHM under the assumption that the LSF is a normalized Gaussian, and one for the UCD RV. We also had fixed parameters for the scaling of the telluric model with airmass, which we define later in this section, as well as the $V\sin{i}$ and effective temperature, $T_{\rm{eff}}$, for each UCD, which we determined separately and fixed in all subsequent analyses. We fit our model to the NIRSPEC data in a least squares sense using an implementation of the AMOEBA \citep{nelder1965,press1986} fitting method to minimize $\chi^{2}$.

\begin{figure}
\plotone{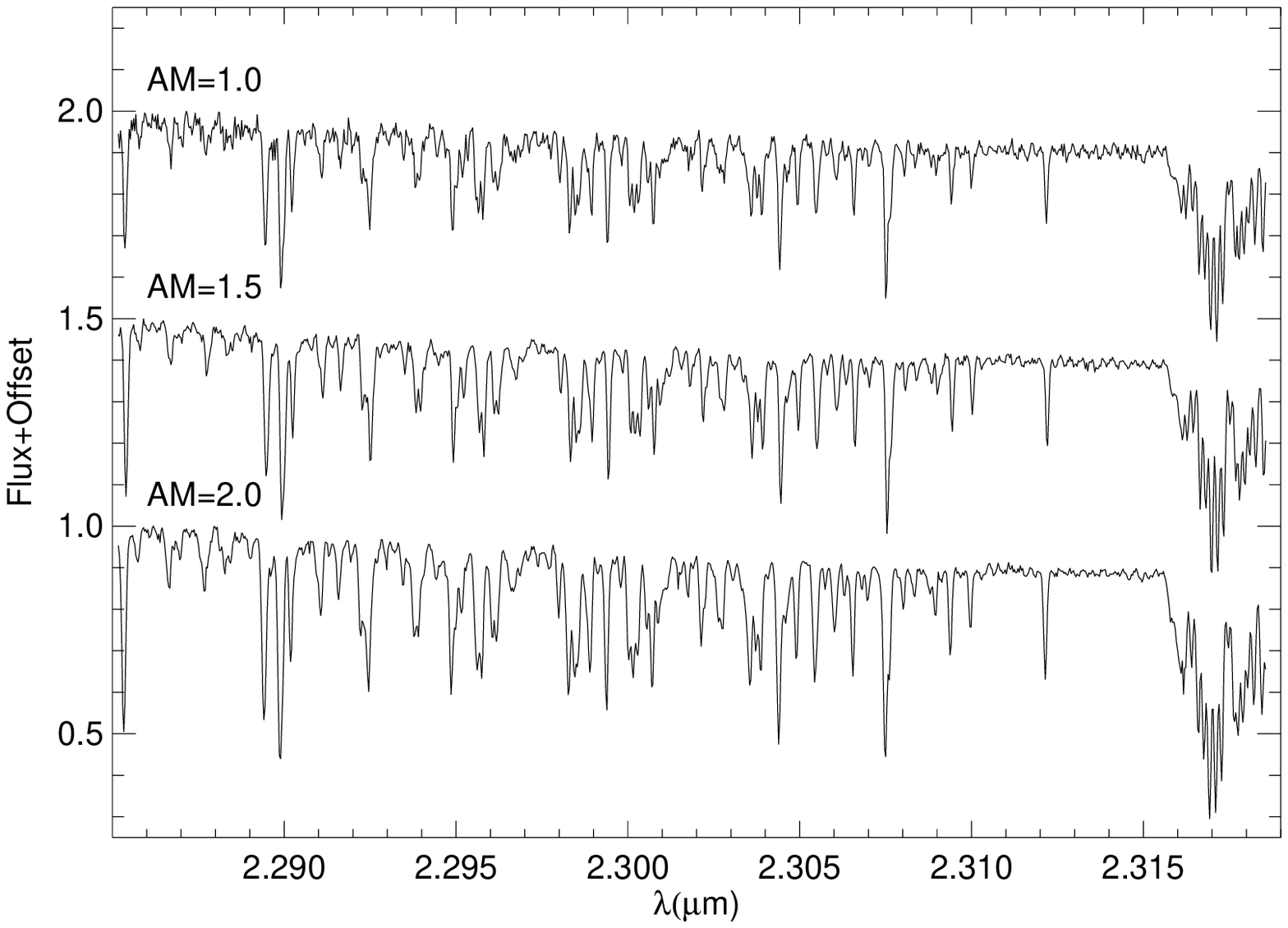}
\caption{Examples of NIRSPEC A star spectra at three different airmasses. These absorption features, due mainly to CH$_{4}$, serve as a simultaneous wavelength reference. The depths of the lines increases as the optical depth of the atmosphere increases with higher airmasses. The ringing between 2.310 and 2.315 $\mu m$ is the fringing pattern discussed in Section 5.1.}\label{astars}
\end{figure}

\begin{figure}
\plotone{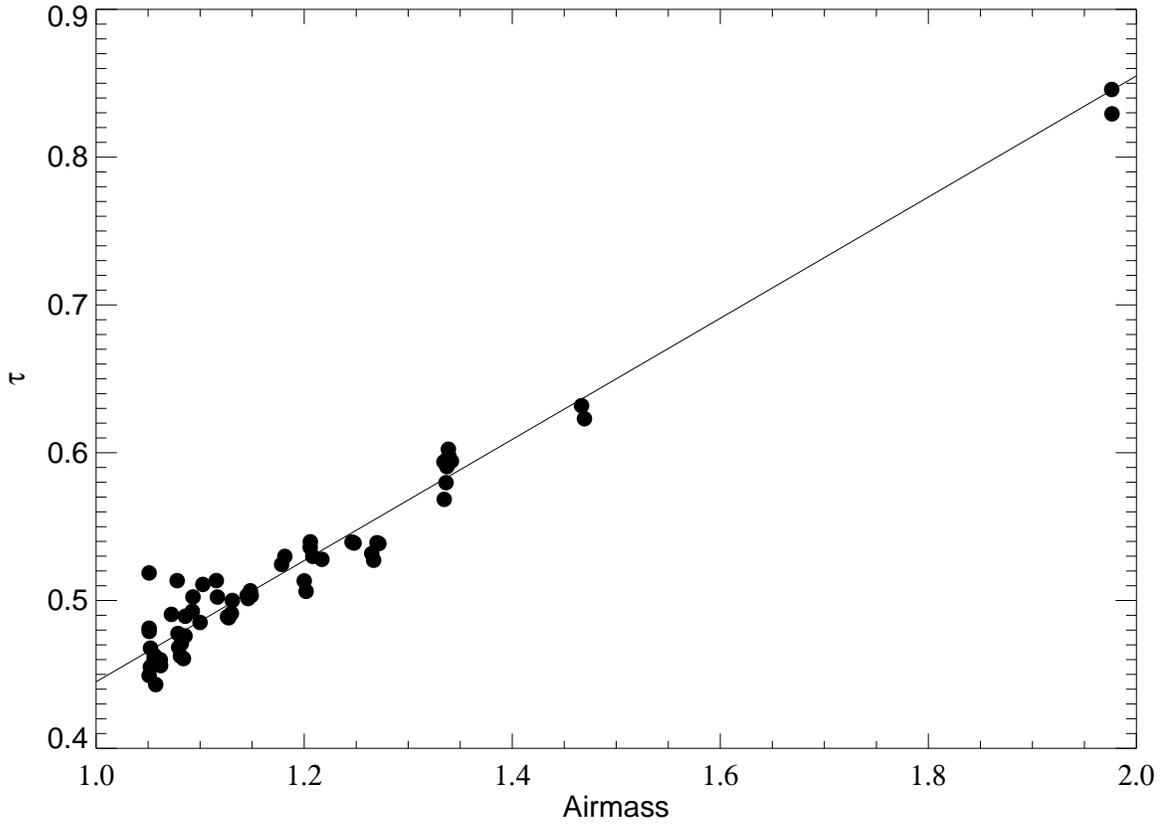}
\caption{Results of fits to A star spectra for the single parameter, $\tau$, that is used to scale the depths of the lines in the telluric template with airmass.}\label{tau}
\end{figure}

We began our analysis of the NIRSPEC data by selecting a training sample of 200 A star observations acquired over the course of the survey at a range of airmasses from 1.0 to 2.0. Examples of A star spectra at a range of airmasses are shown in Figure \ref{astars}. We modeled these observations without including the UCD template in order to refine our fitting procedure, determine the overall distribution of the best-fit wavelength and LSF parameters, and model the changes of the depths of telluric lines with airmass. Starting from the nominal 10 model parameters (no RV) we added an additional free parameter for the scaling of the depths of the telluric lines with airmass. The \citet{livingston1991} data is at airmass (AM) of 1.5, but our data were acquired at a wider range of airmasses. At higher airmass the optical depth of the atmosphere increases and we expect the telluric line depths to increase. We assumed a one parameter scaling of the depths of the telluric lines from the AM=1.5 model ($T_{0}$) with airmass $T=T_{0}^{\tau}$. Based on the fits to the A star training sample, we found that the telluric line depths observed at the summit of Mauna Kea, shown in Figure \ref{tau}, were well fit with $\tau=AM\times 0.432$. After this initial determination of the telluric scaling we adopted this relation for $\tau$ for all of the subsequent fitting of NIRSPEC spectra and fixed this parameter in the modeling of each UCD spectrum. 

\begin{figure}
\plotone{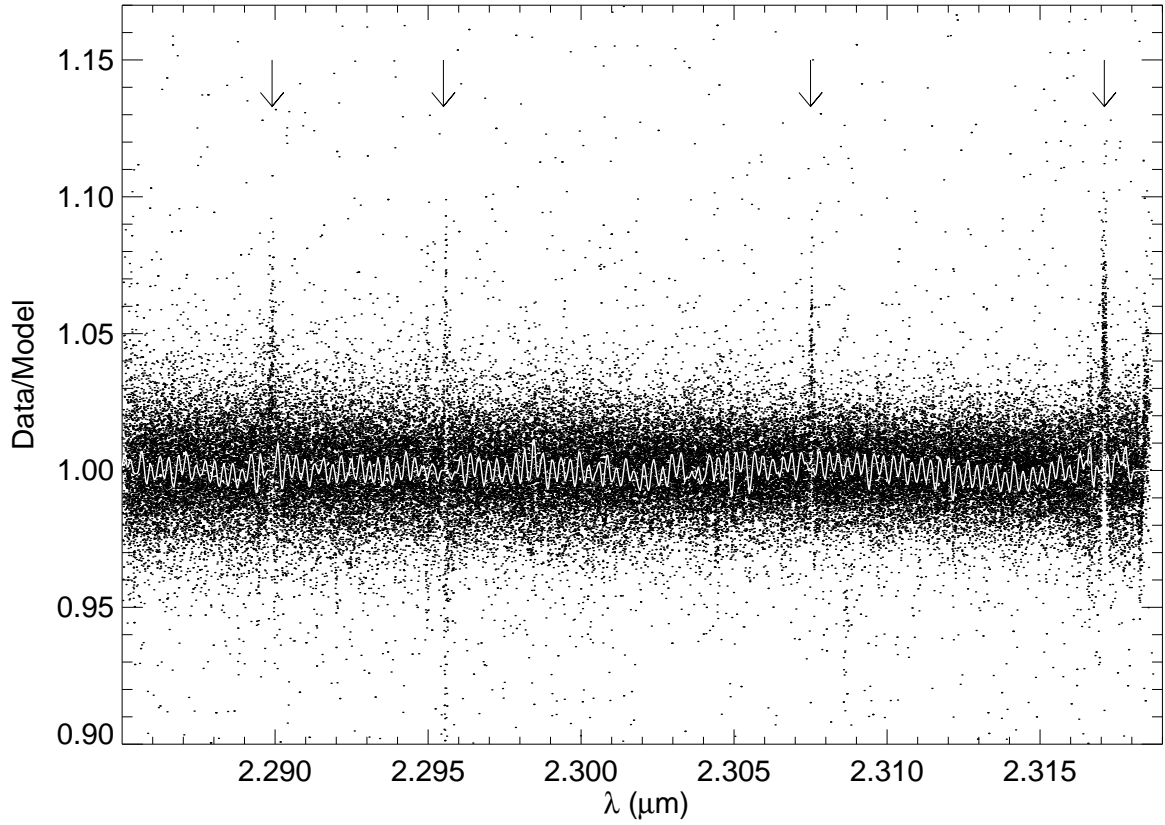}
\caption{Binned residuals of fits to 200 A star spectra. Several lines that do not follow the same depth scaling with airmass are indicated with downward errors.  These lines may be H$_{2}$O absorption features. A small region around each of these features receives no weight during the spectral fitting process. The fringe model described in Section 4 is over plotted in white.}\label{fig2}
\end{figure}

In order for our fitting procedure to successfully converge on the correct model parameters, excellent initial estimates of the parameters were required. The A star training sample allowed us to determine the average model parameters related to the spectrograph and select good initial values for each parameter. We found that our nominal model resulted in A star fits with $\chi_{\nu}^2\sim2-4$. Adding additional parameters for the wavelength solution did not significantly improve the overall quality of the fits. We also found that conducting an initial cross-correlation of the first 200-pixel chunk with a nominal model of a NIRSPEC A star observation allowed for the determination of the zeroth-order term in the wavelength solution to better than 0.1 of a NIRSPEC pixel, sufficiently precise for AMOEBA to reliably converge on the correct wavelength solution. We investigated how the LSF changes across the spectral order by fitting portions of A star spectra independently. While the FWHM of the Gaussian LSFs across the spectra change with time, we found that the ratios of the FWHM of the best-fit Gaussian LSFs in different portions of the spectra were relatively constant so that a single fixed parameter could be used to describe the slow change in the width of the LSF across the order. By looking at the average residuals of the fits to all 200 A stars we also identified individual telluric lines that did not follow the scaling with airmass. In Figure \ref{fig2} we show four lines that are likely not $CH_{4}$ absorption features but may be features due to $H_{2}O$  \citep{rothman2009}. We excluded a small spectral region (ten pixels) around each of these features in all of our fitting of the UCD spectra by assigning zero statistical weight.

The residuals of the model fits to the A star sample are shown in Figure \ref{fig2}. With our instrumental setup a fringe-like modulation was often seen in the extracted spectra and is clearly visible in the A star residual around 2.314~$\mu$m in Figure \ref{astars}. This pattern is likely due to an internal reflection in an optical element near the detector \citep{brown2003} and is generally described as the superposition of sinusoidal patterns with amplitudes of roughly $1\%$ and periods of approximately $0.3$~nm. An additional complication is that the spatial frequency of the fringe pattern is similar to the spatial frequencies of absorption features seen in slowly rotating UCDs, so it can not be easily removed without degrading the signal that we wish to model.
Like \citet{brown2003} we found that the fringing pattern varies somewhat over time in phase, frequency, and amplitude. Assuming that the fringe signal is multiplicative (as opposed to additive) we built a model of the average fringe signal in wavelength space by averaging the A star residuals ($Data/Model$) in bins of width $0.03$~nm. This model, shown in Figure \ref{fig2}, is fixed in wavelength space and has an amplitude of approximately $1\%$ and is quasi-periodic with a dominant period of $0.3$~nm. Including this fringe model in the fitting of the A star sample resulted in sigificant improvements in $\chi^2$ for the highest S/N spectra though had negligible impact on the resulting overall distribution of the best-fit model parameters. Given that the fringe pattern is a quasi-periodic multiplicative modulation it is possible that the average shapes of the telluric lines, and therefore the resulting wavelength solutions, could be biased as a function of the relative phase of the fringe and the telluric absorption features. At the same time, the fact that the fringe model is not strictly periodic could mean that this effect averages out over the spectrum, reducing the impact on the resulting fits. Given the small amplitude of the fringe model compared to the S/N of our A star observations (S/N$\sim200$) it is perhaps not surprising that it is not an important factor in our fits.

We fit the UCD spectra following an iterative process using the average parameter values determined from the A star analysis as the starting parameters for AMOEBA. As with the A star analysis, we used the first 200-pixel chunk of each spectrum to estimate the zeroth-order term of the wavelength solutions by cross correlating against a nominal telluric model. This first 200-pixel chunk (2.585 to 2.592 $\mu$m) of the UCD spectrum is relatively devoid of stellar absorption features and so fitting this chunk to the telluric-only model was useful both for determining 
the starting wavelength position as well as the single LSF FWHM parameter. We also produced an initial RV estimate by cross correlating a spectral region with rich CO features (2.298 to 2.305 $\mu$m) against a fiducial UCD model at zero velocity. This step, which we found necessary for ensuring convergence of the fitting process, provides a rough ($\pm$10 km s$^{-1}$) estimate of the RV that includes shifts due to barycentric motion.
We estimated the $T_{\rm{eff}}$ and $V\sin{i}$ for each UCD in a two step process. We began by fitting each spectrum of each object to a grid of 360 UCD models spanning $1200<T_{\rm{eff}}<2400$~K and $9<V\sin{i}<100$~km$^{-1}$. For each NIRSPEC spectrum we found the global minimum $\chi^{2}$ in the $T_{\rm{eff}}$ and $V\sin{i}$ grid resulting in one estimate of each parameter for each spectrum. Using these initial estimates we set a single $T_{\rm{eff}}$ for each object by selecting the UCD model that produced the lowest average $\chi^2$ values over all of the spectra the object. With $T_{\rm{eff}}$ fixed for each object, we ran a second set of fits over a finer grid in $V\sin{i}$ and then fit for the minimum of the resulting curve of $\chi^2$ as a function of $V\sin{i}$ to estimate the best-fit $V\sin{i}$ for each spectrum. We adopted the simple average of the individual estimates from each spectrum as the fixed $V\sin{i}$ of the UCD and the scatter of those estimates as the error on the $V\sin{i}$. These two parameters, $T_{\rm{eff}}$ and $V\sin{i}$, were then fixed for all subsequent analyses. We determined the lower limit on $V\sin{i}$, set by the resolution of NIRSPEC, by estimating the value below which changes in $V\sin{i}$ did not improve the $\chi^{2}$ of the fits to UCDs that are known to be slow rotators.

\begin{figure}
\plotone{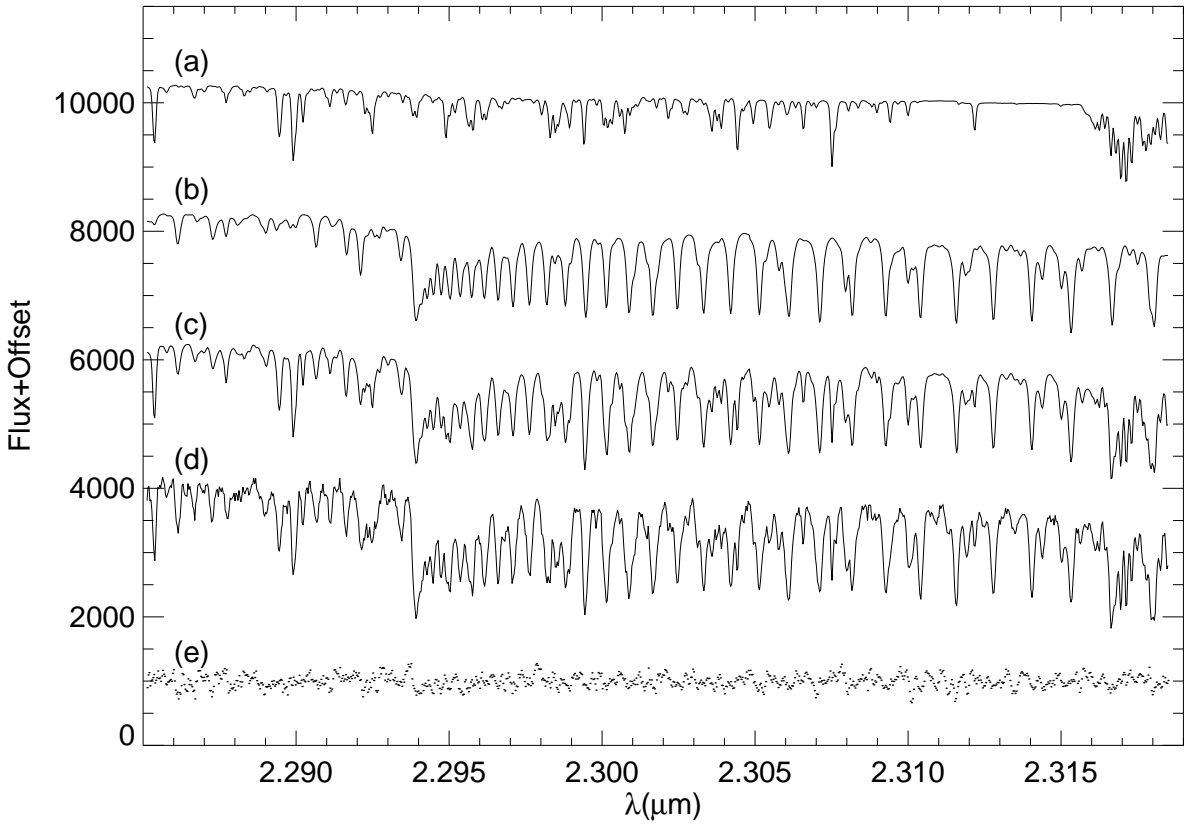}
\caption{Schematic example of the forward modeling process described in Section 4. a) observed telluric spectrum, b) synthetic L dwarf spectrum broadened to account for stellar rotation, c) model from product of (a) and (b) convolved with the spectrograph LSF, d) NIRSPEC observation of 2M1048+01, e) residuals of model fit to data.}\label{panel}
\end{figure}

With the global UCD parameters fixed ($V\sin{i}$ and $T_{\rm{eff}}$) for each target, we estimated the RV of each observation of each target using the same fitting process. Each spectrum was fit in two stages, rejecting outliers following an initial fit and then fitting again using the best fit parameters from the first fit as the new starting parameters. We conducted extensive tests using artificial spectra to ensure that the correct minima were being found by AMOEBA by conducting fits with fixed wavelength solutions over a large, high-resolution parameter space. We generated the artificial spectra based on the wavelength solutions found in the A star analysis, the synthetic UCD templates over a range of $T_{\rm{eff}}$ and $V\sin{i}$, and noise properties representative of the actual NIRSPEC data. We found that with good starting values our AMOEBA fitting procedure reliably converged on the true
minimum of $\chi^{2}$ and the correct model parameters. The formal reduced $\chi^{2}$ of the UCD fits fall in a wide range (2$<\chi_{\nu}^{2}<8$). Unlike with the A star analysis, where the $\chi^{2}$ is likely dominated by the noise
properties of the NIRSPEC detector and limitations in our LSF model, the UCD fits may be dominated by the mismatch between the theoretical stellar templates and the UCD and we expect some larger values of $\chi^{2}$.  For example, systematic discrepancies are sometimes seen in the structure of the CO bandhead or in the depths of individual CO features
longward of the bandhead. We emphasize that systematic deviations between the spectra and the theoretical models may not be shortcomings of the models themselves but rather a symptom of the small range of our library of synthetic templates in terms of    $\log{g}$, $f_{\rm sed}$, and metallicity. An example of a NIRSPEC spectrum, the best fit model, and the residuals is shown in Figure \ref{panel} and the top panel of Figure \ref{empiric}. 

We estimated the statistical uncertainty on each RV measurement, $\sigma_{RV}$, based on the photon-limited Doppler precision (PLDP) presented in \citet{butler1996}

\begin{equation}
\sigma_{RV}=\left[\Sigma \left(\frac{dM/dRV}{\epsilon_{D}}\right)^{2}\right]^{-1/2}
\end{equation}
where the sum is over all pixels, $dM/dRV$ is the rate of change of the model flux at a given pixel, in velocity units, and $\epsilon_{D}$ is a fractional noise term. Here, $M$ is the UCD (or telluric) component of the best-fit model for each spectrum. In the case of photon noise alone $\epsilon_{D}=\sqrt{D\times G}/{(D\times G)}$ where $D$ is the data in ADU and $G$ is the detector gain. We estimated the PLDP of the best fit model by evaluating the telluric and UCD components of the model separately and adding the two error estimates in quadrature. This accommodates the case of a high S/N observation of a rapidly rotating object where the wavelength solution may be determined precisely from the deep telluric lines but the RV is only poorly determined from the broad stellar features. For the typical S/N of our UCD observations we estimate that the telluric lines themselves limit the RV precision to $\sigma_{RV}>60$ m s$^{-1}$. Our data have significant sources of noise beyond just photon noise. Instead of assuming photon noise alone, we estimated the noise term $\epsilon_D$ from the residuals of the best fit model [$S=D-M$]. The residuals could be dominated by systematic discrepancies between the model and the data, which could result in an over estimation of the noise term based the RMS of $S$ alone. To remove such systematic residuals we first smoothed the residuals [$S'=SMOOTH(S)$] with a boxcar filter of width five pixels and then estimated the noise term 

\begin{equation}
\epsilon_{D}=\frac{\sqrt{\left<(S-S')^{2}\right>-\left<S-S'\right>^{2}}}{D}
\end{equation}
We applied barycentric corrections to the individual RV estimates calculated using the code \textit{bcvcorr} (G. Torres; private communication) and then combined observations (generally nod positions A and B) from the same epoch using a weighted mean.

\section{Radial Velocity Precision}

\begin{figure}
\plotone{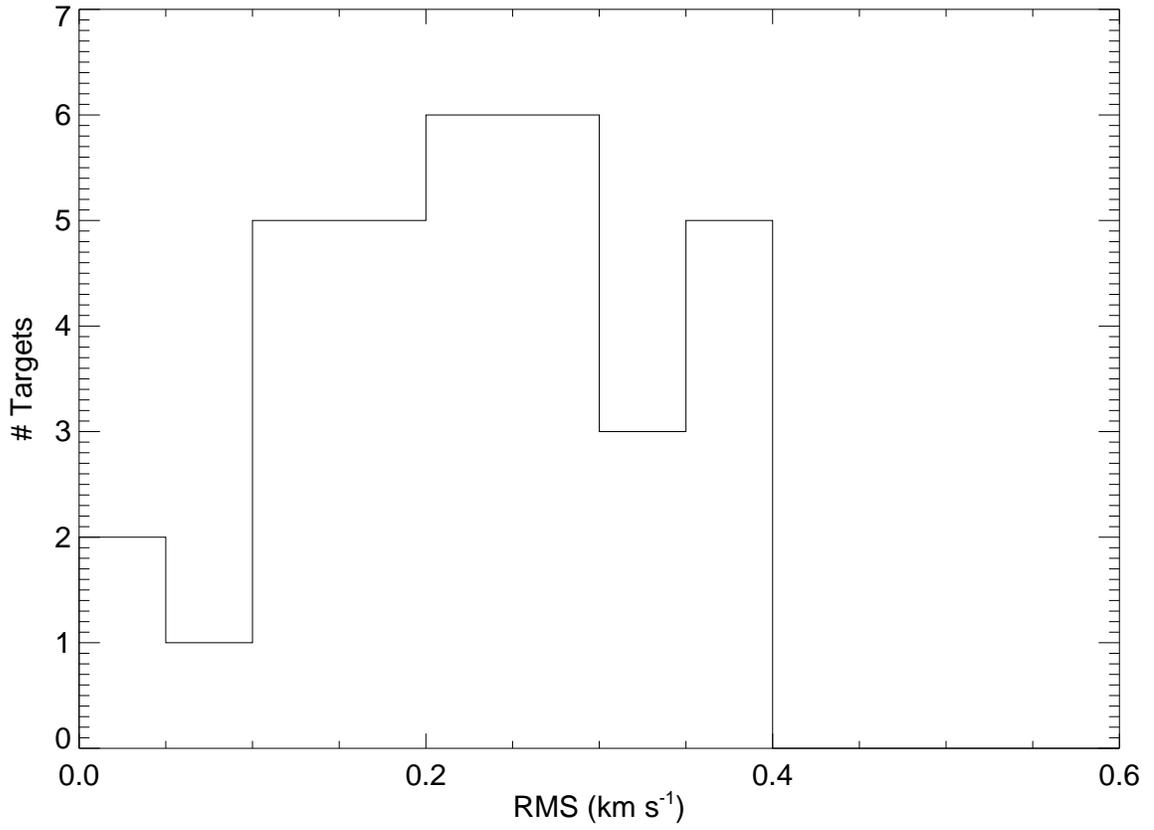}
\caption{Histogram of the measured RV RMS dispersion of the UCDs with observations on three or more epochs and $V\sin{i}<30$ km s$^{-1}$. For the majority of our targets we achieve an overall precision of 100-300 m s$^{-1}$.}\label{rmshist}
\end{figure}

\begin{figure}
\plotone{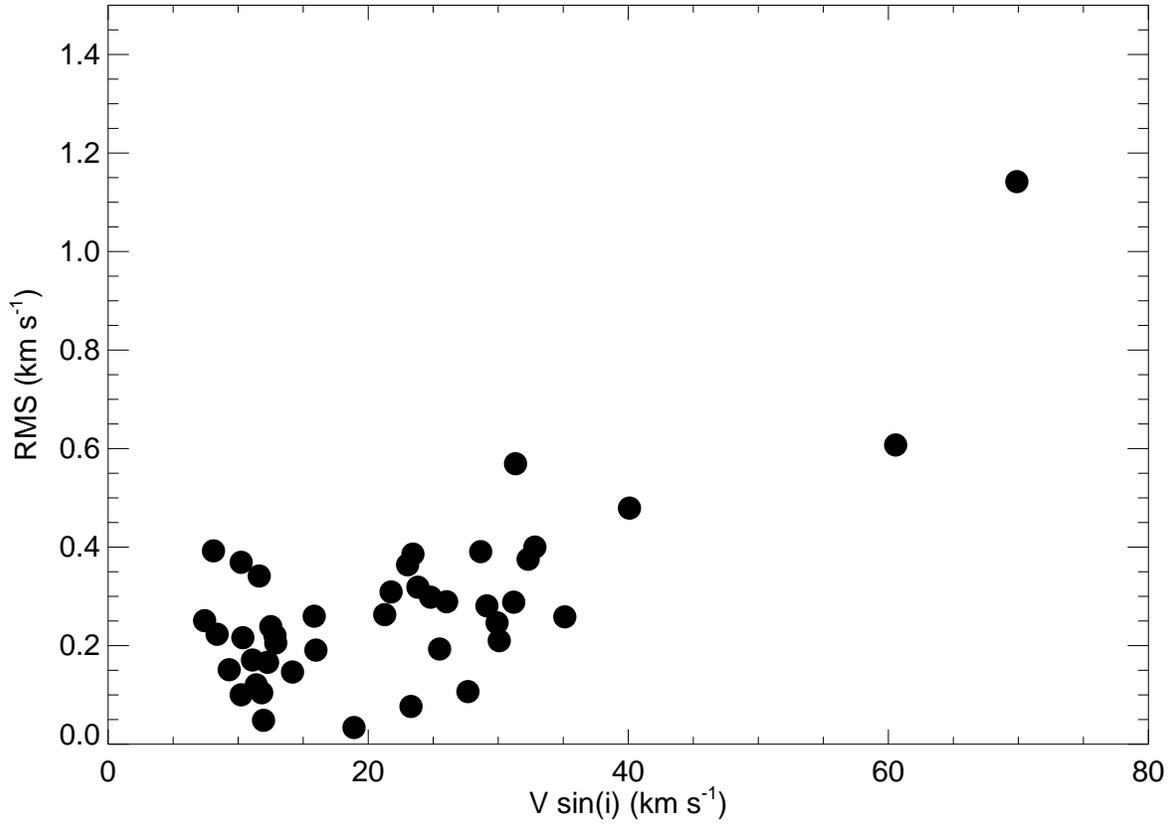}
\caption{Estimated RV RMS dispersions for targets with observations on three or more epochs. Rapid rotation broadens the spectral features of the L dwarfs, decreasing the expected RV precision. We see RMS below 200 m s$^{-1 }$ in a number of our slowly rotating targets. We estimate that the minimum V $\sin{i}$ measurable with our NIRSPEC data is 9 km s$^{-1}$.}\label{vsinrms}
\end{figure}

\begin{figure}
\plotone{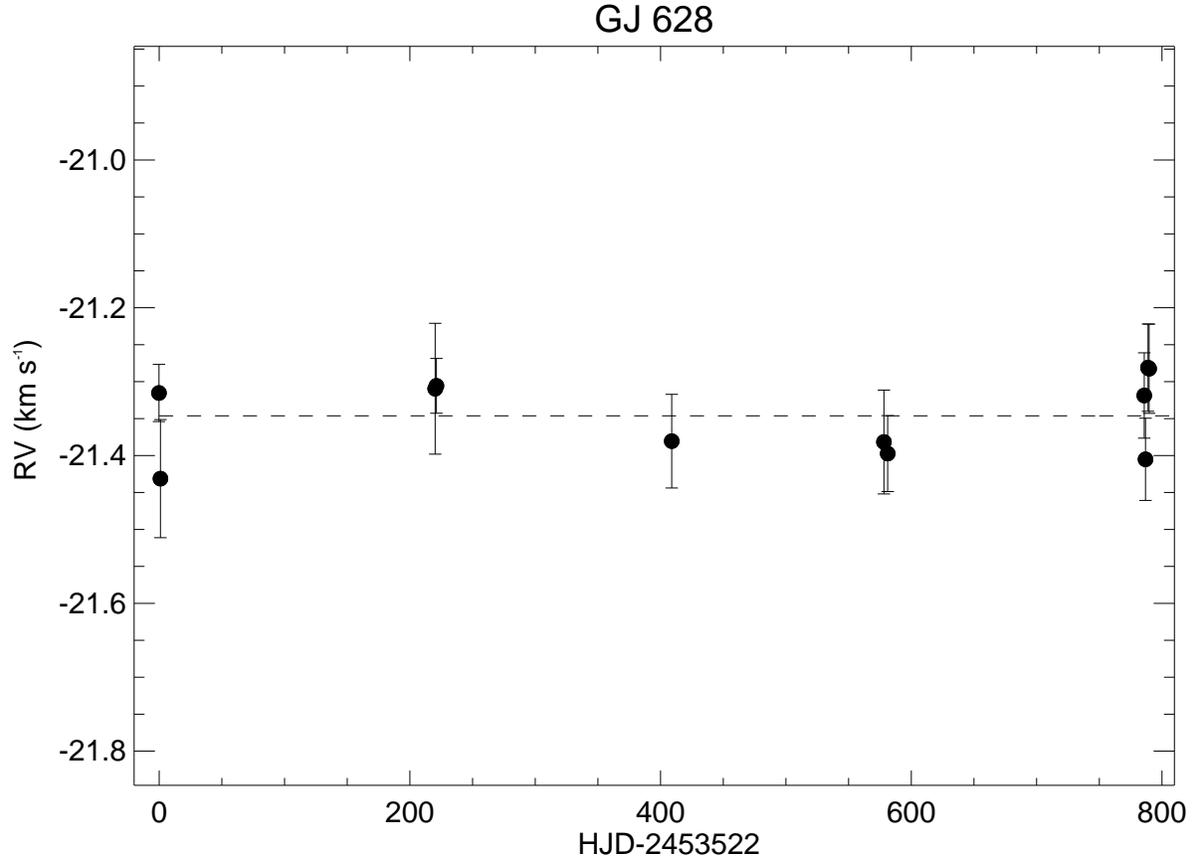}
\caption{RV measurements of the bright, slowly rotating M dwarf GJ 628. We estimate the RMS to be 50 m s$^{-1}$ over the 800 day span of the observations. The scatter in these measurements is fully consistent with the photon-limited Doppler precision error estimate given by Eqn. 2}\label{gj628}
\end{figure}

Based on the standard deviations of the RV measurements of 43 of our targets with observations on three or more epochs, $V\sin{i}<30$ km s$^{-1}$, and excluding known or suspected variables, shown in Figure \ref{rmshist}, we estimate our RV precision to be approximately 100-300 m s$^{-1}$ for slowly rotating UCDs.  In Figure \ref{vsinrms} we compare the measured standard deviation of the RVs of each of our targets to the estimated $V\sin{i}$. As expected, our RV precision degrades significantly for rapidly rotating L dwarfs. During the course of our survey we also observed the bright (K=5.08), slowly rotating (V$\sin{i}<8$ km s$^{-1}$) M3 dwarf GJ 628 as an RV standard. This object is more massive than the stars in our UCD sample, falling outside the $T_{\rm{eff}}$ range of our synthetic templates. For the spectral fitting we used a synthetically generated M dwarf template with $T_{\rm{eff}}=3400$K, $\log{g}=4.8$ computed from updated and improved NextGen \citep{hauschildt1999} models (T. Barman, priv. comm.) From our analysis of 11 epochs of observations of GJ 628 we found an RV RMS of 50 m s$^{-1}$ over a period of 800 days, as shown in Figure \ref{gj628}. We note that the scatter of these RV measurements is well-described by the PLDP error estimates obtained using the technique described in Section 4 [$P(\chi_{RV}^{2}\le)=0.47$] indicating that at least for bright, slowly rotating objects we are achieving an RV precision very close to the photon limit. 

\begin{figure}
\plotone{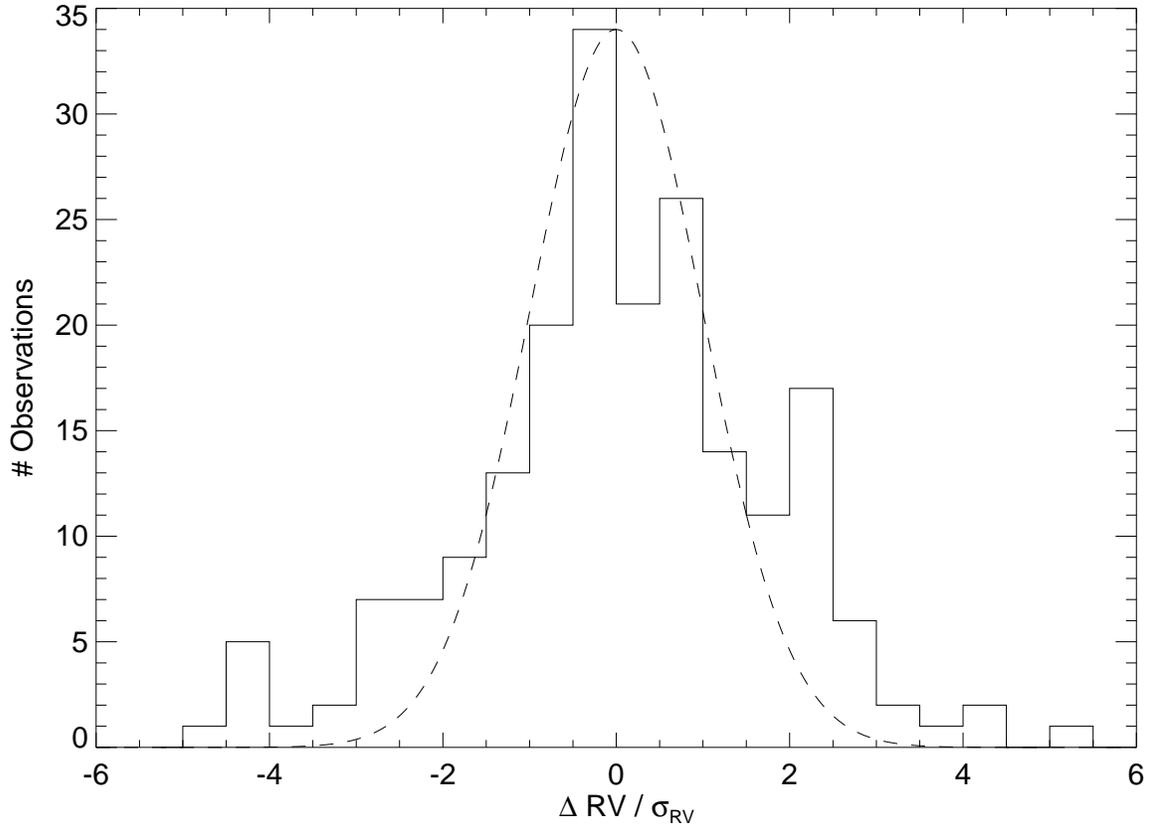}
\caption{Overall distribution of $\Delta RV$ normalized by the PLDP error $\sigma_{RV}$. Here we have assumed that all of the objects in this sub-sample have no intrinsic RV variations. The dashed line is an N[0,1] distribution, the expected distribution if the error estimates correctly describe the data. We scale the empirical error estimates, $\sigma_{RV}$, by a factor of 1.9 in order to make them consistent with the expected N[0,1] distribution.}\label{residnorm}
\end{figure}

\begin{figure}
\plotone{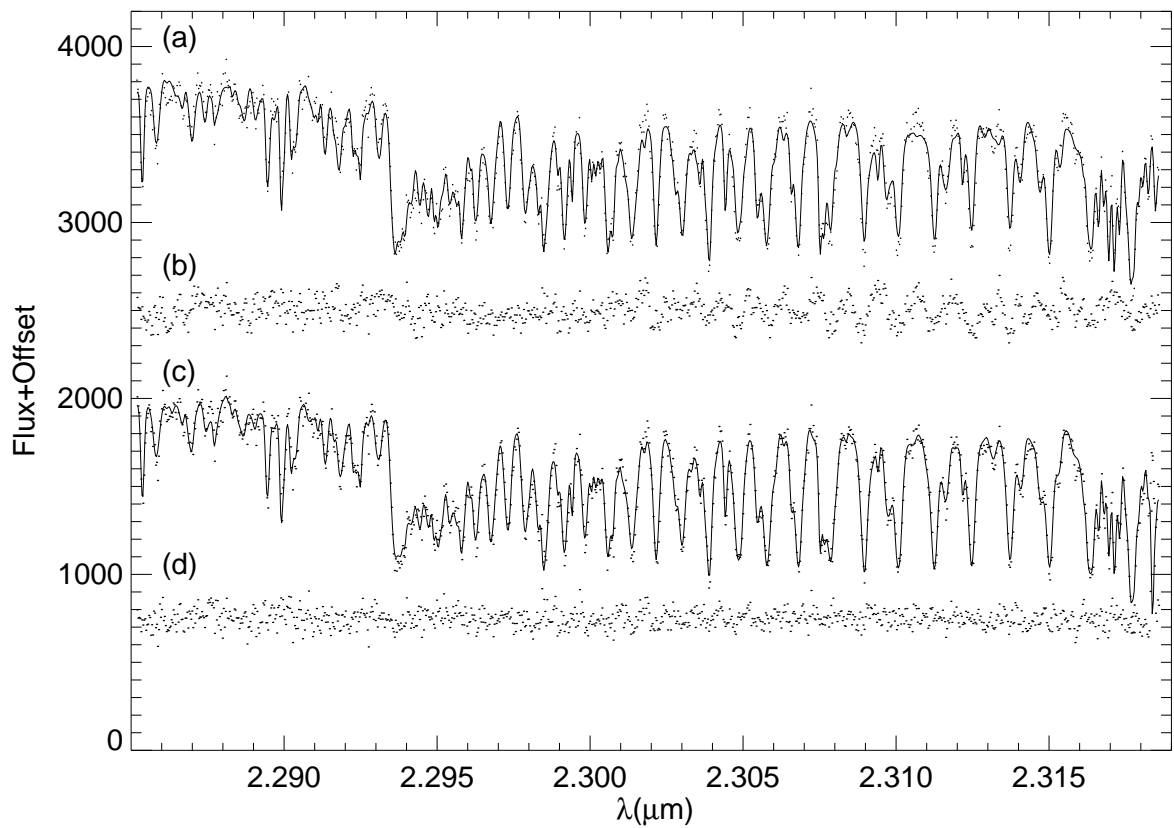}
\caption{Comparison between fits for 2M0835$-$08 with theoretical and empirical templates. a) NIRSPEC observation (points) with best fit model based on theoretical template (solid line). b) residuals of this fit. c) NIRSPEC observation (points) with best fit model based on empirical template described in Section 5.1 (solid line), d) residuals of this fit. The empirical template significantly improves the overall quality of the fit ($\Delta \chi^{2}=420$) but in our tests the empirical templates did not result in an overall improvement in RV precision.}\label{empiric}
\end{figure}

To reliably estimate the statistical significance of any RV variations we detect in our UCD sample it is necessary to understand the underlying errors, both systematic and statistical, on our individual measurements. We have estimated the statistical errors on our individual RV measurements, $\sigma_{RV}$, directly from the data, but these estimates do not take into account systematic effects that may occur between epochs. To investigate the overall statistical properties of our measurements we selected a sub-sample of objects that have observations on three or more epochs and are not known or suspected binaries. Assuming the null hypothesis that each of these UCDs has a constant RV, using the empirically-determined internal errors, $\sigma_{RV}$, for each of the 207 measured RVs we calculated  $\chi_{RV}^{2}=591$ for $207-43=164$ degrees of freedom (assuming one parameter for the constant RV of each target). If we have excluded all of the actual RV variables from this analysis, then this value of $\chi_{RV}^{2}$ indicates that our statistical error estimates are too small and that there is a significant systematic contribution to be included in our overall error model. The overall error distribution for this sub-sample is shown in Figure \ref{residnorm}, which shows that there are significant non-Gaussian tails at $|\Delta$RV$|>2\sigma$. We scaled all of the PLDP errors estimates for the UCDs by a factor of $\sqrt{\chi_{\nu}^2}=1.9$, resulting in a reduction of the $\chi_{RV}^{2}$ to 163.6 for the same number of degrees of freedom $\left[P \left(\chi_{RV}^{2}\le 163.6\right)=0.50\right]$. In some cases it is clear that the scaled errors are too large. For example, the scatter about the best fit orbital solution for the binary 2M0320$-$04, described in Section 6.1, is 135 m $^{-1}$ while the smallest error estimate for a single point in the fit is 178 m s$^{-1}$. Despite this possible overestimation, we used the scaled RV error estimates for all subsequent analyses of the UCD RV measurements. Based on our observations of the RV standard GJ 628 we conclude  that our modeling process can produce RV measurements near the photon-limit over long timescales and that the worse overall RV precision obtained for the UCDs could be due to a number of factors. The tails of the distribution of $\Delta RV$ could be the result of real RV variability since we have only excluded known and suspected binaries from our sub-sample. While the theoretical template used to model GJ 628 is a very good match for its spectrum, mismatch between the theoretical templates and the actual spectra of the UCDs could lead to worse RV precision. Lastly, it is possible that the much longer integration times used for the UCD observations compared to the GJ 628 observations lead to systematic effects that are not encompassed in our model. 

\subsection{Additional Tests}

In an effort to increase the overall precision of our RV measurements, as well as to address some technical issues that may be important for efforts to achieve high RV precision in the NIR with NIRSPEC or similar instruments, we experimented with some modification to the standard fitting process described in Section 4. The first was the inclusion of the fringe model derived from observations of A stars into the UCD fits. Including this fringing model in the A star fitting process led to significant improvements to the resulting $\chi^2$, though had negligible impact on the overall statistical properties of the resulting model parameters. Similarly, including the fringe model in the analysis of the GJ 628 observations did not result in a statistically significant decrease in the resulting RV scatter. While the A star and GJ 628 observations typically have S/N$\sim200$, the UCD observations have $50<$S/N$<100$. The low-amplitude flux modulation of the fringe pattern is small compared to the read noise and photon noise and so was not expected to significantly impact the fitting process. We did experiment with including the fringe model in the flitting of the UCD spectra and found that the overall statistical properties of the resulting RV measurements were consistent with or without the fringe model and that the photon-limited errors still needed to be scaled by a factor of $1.9$ to account for the observed scatter in the RV measurements of individual objects. The fringe model was not included in the final UCD RV results presented here.

While our theoretical templates are generally a very good match for the UCD spectra, for some individual objects there are significant systematic discrepancies in the shapes of the spectral features.  This is most likely due to the fact we are only fitting a small library of synthetic templates to our spectra and a wider range of $\log{g}$, $f_{\rm sed}$, metallicity values,  and line damping treatments could significantly improve the fits for some objects.
If we have multiple observations of an object, and we can assume that the object has a constant RV, then it is possible to build an empirical spectral template from the observations themselves. We began by fitting the spectra using the theoretical templates following the standard procedure described in Section 4. Assuming that the wavelength solutions were sufficiently well-determined by this initial fit we divide the data by the telluric component of our best fit model, resulting in a normalized UCD spectrum that is free from atmospheric absorption features. We corrected the wavelength solutions for the known barycentric velocity of each spectrum and then averaged all of the normalized spectra in 0.01~nm bins to create an empirical spectral template. For a slowly rotating UCD, where $V\sin{i}$ is comparable to or smaller than the spectrograph resolution, it is necessary to account for the broadening of the stellar spectral features by the LSF of the spectrograph. To do this we assumed a Gaussian LSF with FWHM=$0.076$~nm and carried out an iterative deconvolution following \citet{lucy1974} on the empirical template in an attempt to recover the un-broadened spectrum. Using this method we created an empirical template for the rapidly rotating ($V\sin{i}=30.1$ km s$^{-1}$) L dwarf 2M0652+47 and found that the overall fit residuals were reduced to the level of $1\%$. We also generated templates for more slowly rotating objects and a comparison between a theoretical template and an empirical template for the slowly rotating L dwarf 2M0835$-$08 is shown in Figure \ref{empiric}. While this technique did in some cases result in templates that were significantly better fits to the observed spectra, there are a number of drawbacks. The first is that a large number of spectra are required to produce the empirical template, preferably more than five, in order to robustly average in wavelength bins. The second is that we began by assuming that the object has a constant RV. Intrinsic Doppler shifts will result in a broadened spectral template and, particularly if the number of spectra is small, the empirical template may result in biased RV estimates. We found that using empirical templates to fit the subset of our sample with observations on five or more epochs did not yield RVs with a smaller dispersion. With our current data the creation of empirical templates may be most useful when looking for temporal changes in the residuals of the fits, which could be evidence for a faint companion.

As the precision of RV  measurements at optical wavelengths has improved to the level of 1 m s$^{-1}$ it has become clear that modeling the shape of the spectrograph's LSF is critically important. \citet{butler1996} used a basis set of 11 Gaussian functions to model the subtle changes in the LSF asymmetries in their $I_{2}$ cell observations. Spectrograph LSF asymmetries may be inherent to the instrument or, in a slit spectrograph, they may arise from guiding errors inducing variations in the stellar position along the slit. We modeled the NIRSPEC LSF as a symmetric, Gaussian function that has a linear variation in width across the order. We experimented extensively with using multiple Gaussians to describe a more complex LSF in a manner similar to \citet{butler1996}. Owing to the limited S/N of our data, we could not reliably model the LSF asymmetry in either the A star or UCD spectra. Future improvements in the RV precision obtained using this technique will require a detailed modeling of the spectrograph LSF but the current spectral resolution and S/N may not motivate such an analysis.

\section{Radial Velocity Variables}

The measurements from our NIRSPEC survey, listed in Table 2, represent the largest available sample of high-resolution, high-S/N, NIR observations of L dwarfs. The primary goal of this survey has been the detection of RV variations due to unseen companions. With an overall RV precision of approximately $200$ m s$^{-1}$ for slowly-rotating UCDs, we are sensitive to many types of binary systems. The RV semi-amplitude, $K_{1}$, of the primary (more luminous) star in a spectroscopic binary (or planetary system) is

\begin{equation}
K_{1}  \approx 654\ {\rm{m}\ \rm{s}^{-1}} \left(P\over{3\rm{d}}\right)^{-1/3} \left({M_{\rm{2}}}\over{M_{\rm{J}}}\right)  \left({M_{1}+M_{2}}\over{0.1M_{\sun}}\right)^{-2/3}  \left(1-e^{2}\right)^{-1/2}  \sin{i}
\end{equation}
The targets in our sample are expected to have masses from near $0.1M_{\odot}$ down to the late-L dwarfs at $0.05 M_{\sun}$ \citep{burrows2001}. If our survey is sensitive to signals with $\Delta$RV$>1$km s$^{-1}$, then we may detect systems ranging from equal-mass binaries with periods of up to a decade to giant planetary companions with orbital periods of several weeks or less. We expect that the majority of our targets are not spectroscopic binaries \citep{allen2007} and therefore will not exhibit RV variations. With the scaling of our PLDP error estimates our RV measurements are fully compatible with this hypothesis. The calculated $\chi_{RV}^{2}$ for each UCD, assuming constant RV, and corresponding probabilities are given in Table 3. We identified five with large $\chi_{RV}^{2}$ having probabilities $P(\chi_{RV}^{2}\le)$, the statistical probability of getting a smaller value of 
$\chi_{RV}^{2}$, greater than 0.999. We note that three of these systems exhibiting significantly large $\chi_{RV}^{2}$ are either known or candidate binaries. Two of these are known binaries already in the literature (2M0746+20, LSR1610$-$0040), one is a new binary identified in the early stages of this survey (2M0320$-$04), and one target exhibits a possible long-term RV trend (2M1507$-$16). LSR0602+39 exhibits statistically significant RV variations that show no evidence for periodicity or a long-term trend. This object is an L subdwarf discovered in the galactic plane by \citet{salim2003} and, based on the detection of Li in the spectrum, is thought to be a brown dwarf. Possibly because of its low metallicity, our theoretical spectral templates are a comparatively poor fit for the observed spectrum of this object, particularly around the CO bandhead, and as a result our theoretical error estimates are possibly underestimated. We also compared our RVs to the few that exist in the literature, including those from \citet{blake2007a}, to look for any additional evidence of long-term trends. While we found overall good agreement with our measurements we note that the RV for Kelu-1 (2M1305$-$25) reported by \citet{basri2000} from observations in 1997 differs by about 10 km s$^{-1}$ from our measurement in 2003. Kelu-1 is a binary, or possibly triple, system with an estimated orbital period of 38 years \citep{gelino2006,stumpf2008}, with an expected RV semi-amplitude of 3-4 km s$^{-1}$. We found no such offset for the other objects observed in common with \citet{basri2000}.

Measuring the reflex motion of the primary component of a single-lined binary (SB1) system, $K_{1}$, allows us to measure the mass function, a transcendental equation that involves $M_{1}$, $M_{2}$, and $\sin{i}$
\begin{equation}
f\left(m\right) = \frac{M_{2}^{3}\sin^3{i}}{\left(M_{1}+M_{2}\right)^{2}} =\left(1.0361\times10^{-7}\right)\left(1-e^{2}\right)^{3/2}K_{1}^{3}P M_{\sun}
\end{equation}
where $P$ is in days and $K_{1}$ is in km s$^{-1}$. In a double-lined spectroscopic binary system (SB2) the measurements of $K_{1}$ and $K_{2}$ can be combined to directly estimate the mass ratio $q=M_{2}/M_{1}$. Additional observations, such as astrometric observations of the binary orbit, are required to determine $\sin{i}$ and to measure any of these quantities independently. We used a non-linear Levenberg-Marquardt fitting scheme \citep{markwardt2009} to fit the six Keplerian orbital parameters to our data and analyze the RV variations of four of our targets. The six orbital parameters we fit are: $K_{1}$, the RV semi-amplitude of the primary; $\gamma$, the systemic velocity of the center of mass of the system; $\omega$, the longitude of periastron; $e$, the eccentricity; $T_{0}$, the time of periastron passage; $P$, the orbital period.

\subsection{2M0320$-$04}

\begin{figure}
\plotone{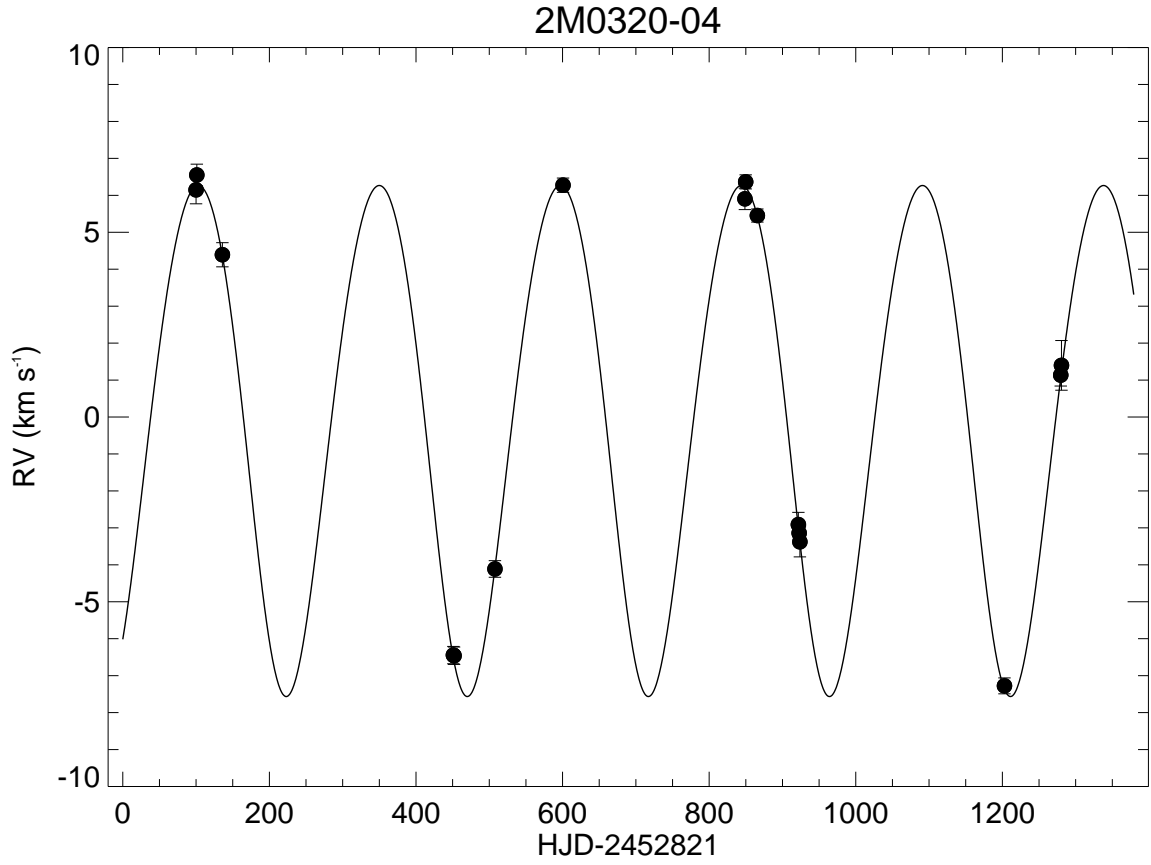}
\caption{Best fit orbital solution for 2M0320$-$04. The scatter of the data about the model is $135$~m s$^{-1}$.}\label{0320}
\end{figure}

We discovered the SB1 spectroscopic binary system 2M0320$-$04 early in our survey \citep{blake2008b} and its binarity was suggested independently using spectral fitting methods by \citet{burgasser2008}. With a relatively short period ($P=246.9\pm0.52$~d) and large RV semi-amplitude ($K=6.92\pm0.12$~km s$^{-1}$), this system was easily detected in our NIRSPEC data and was observed on 16 epochs during the course of our survey. We fit the spectroscopic orbit using RV measurements that are improved over those in \citet{blake2008b} and found consistent results with a scatter about the fit of $135$~m s$^{-1}$, improved from the $350$ m s$^{-1}$ scatter found using the analysis pipeline described in \citet{blake2008b}. The new RV measurements and the orbital solution are shown in Tables 4 and 5 and Figure \ref{0320}. 
Based on the spectral analysis of \citet{burgasser2008}, this system is thought to be composed of a late-M dwarf and an early-T dwarf with masses $M_{1}\simeq0.08M_{\sun}$ and $M_{2}\simeq 0.054M_{\sun}$. As a single-lined system, our spectroscopic orbit results in a mass function and without additional observations to determine $\sin{i}$, the individual masses, or their ratios, can't be directly measured. While this system is expected to be too narrowly separated ($\theta\sim$17 mas) to be resolved with adaptive optics systems on current telescopes,  with  more sensitive observations the spectral lines of the secondary should be detected. In that case, the ratio of the masses and the ratio of the brightnesses can be determined using a technique like TODCOR \citep{mazeh1994}. Such measurements would allow for some of the first direct tests of coeval UCDs at field ($t>1$Gyr) ages, providing important constraints on theoretical models. The flux ratio at $K$-band, where our NIRSPEC data were taken, is expected to be between $100:1$ and $5:1$, and thus may be below our detection limit. The estimated mass ratio of $0.6\lesssim M_{2}/M_{1} \lesssim 0.8$ \citep{burgasser2009} would indicate a maximum velocity separation between the two components of $15\lesssim \Delta V \lesssim18$ km s$^{-1}$, potentially resolvable with NIRSPEC.

In order to search for the spectral lines of the secondary we created an empirical spectral template for 2M0320$-$04 following the same method described in Section 5.1 with the additional step of correcting the wavelength solutions for the SB1 orbit determined above. We found no evidence for temporal variations in the residuals of the fits. Using an empirical template to fit the NIRSPEC observations of the rapidly rotating ($V\sin{i} =30.1$ km s$^{-1}$) object 2M0652+47 we found that the residuals to the fits were at the level of $1\%$. For rapidly rotating objects any changes in the spectrograph LSF have a small impact on the empirical template generation. This is not the case for 2M0320$-$04, where $V\sin{i}=16.7$ km s$^{-1}$ is slightly larger than the width of the LSF. In order to quantify the impact of this effect we studied fits with an empirical template for the NIRSPEC observaitions of 2M0835$-$08 ($V\sin{i}=14.18$ km s$^{-1}$). In this case we found fit residuals using the empirical template were as large as $5\%$ in the cores of CO features. In the future it may be possible to create an empirical template by deconvolving the individual spectra with a more realistic estimate of the LSF, increasing our sensitivity to faint companions.

 \subsection{LSR1610$-$0040}

\begin{figure}
\plotone{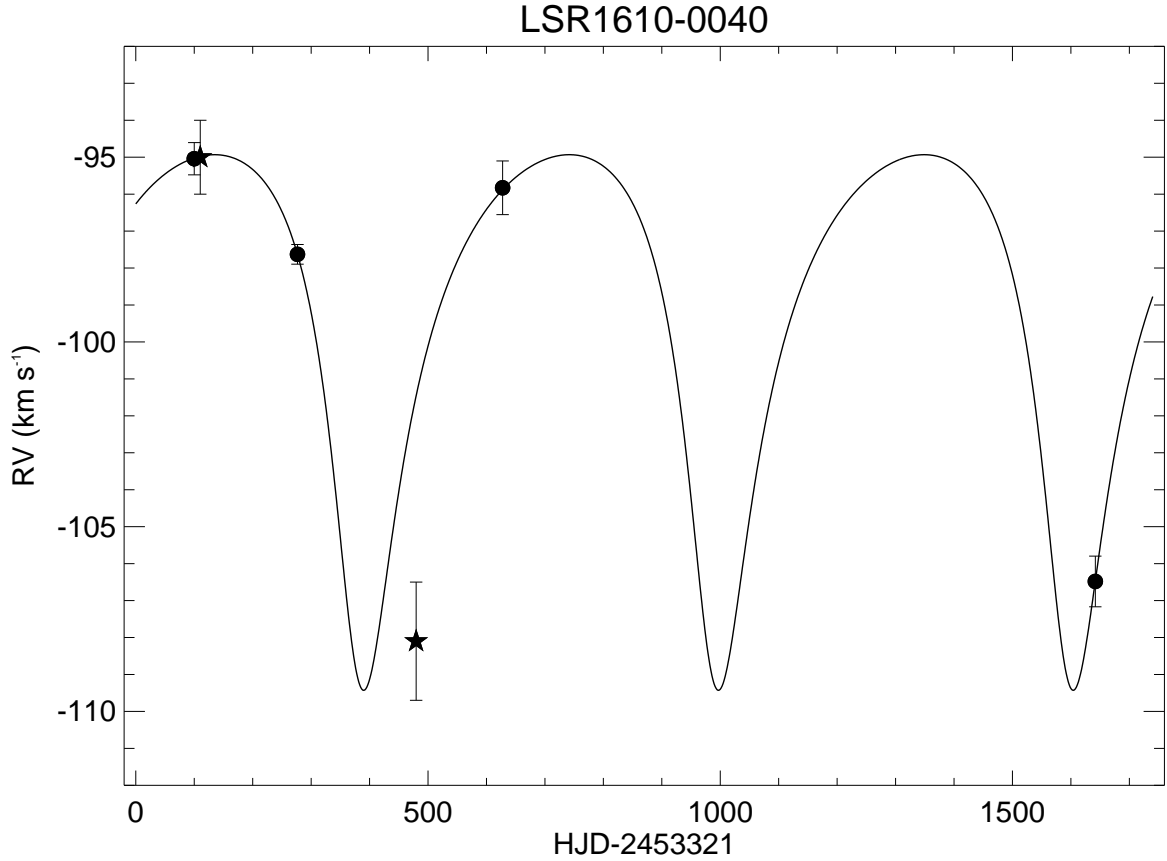}
\caption{Orbital solution for the subdwarf binary LSR1610$-$0040 (solid line) assuming $P$,$e$, and $\omega$ from \citet{dahn2008} and fitting for $\gamma$, $T_{0}$, and $K_{1}$. Circles are our measurements, stars are from the literature \citep{dahn2008,basri2006}. The single point from \citet{dahn2008} is incompatible with our RV measurements and the astrometric orbital solution and is excluded from our analysis.}\label{lsr16}
\end{figure}

LSR1610$-$0040 (LSR1610) is a known binary with a low-mass primary and an orbital solution based on fits to the astrometric motion of the center of light of the system \citep{dahn2008}. We combined our four measurements with two others from the literature \citep{reiners2006, dahn2008} and found good agreement between our RV measurements and the orbital parameters determined from the astrometric orbit alone. The single measurements from \citet{dahn2008} derived from fitting individual atomic resonance lines in the optical spectrum deviates significantly from the expected spectroscopic orbit and is excluded from the following analysis. The single measurement from \citet{reiners2006}, also at optical wavelengths, is nearly coincident with one of our RV measurements and the two agree within 1$\sigma$. We performed a Monte Carlo simulation to estimate the parameters of the spectroscopic orbit by varying $P$,$e$, $i$, and  $\omega$ according to the quoted errors in \citet{dahn2008} while also varying the RV measurements according to their quoted errors and fitting for $K_{1}$, $T_{0}$, and $\gamma$. The RV measurements and best fit spectroscopic orbital orbital parameters are given in Tables 6 and 7 and the orbital solution is shown in Figure \ref{lsr16}. The primary of the system is expected to be very low in mass and assuming that the flux of the secondary is negligible \citet{dahn2008} estimate $0.09<M_{1}<0.098$ M$_\odot$ using the Mass-Luminosity Relation (MLR) from \citet{delfosse2000}. 

The orbital motion of the center of light from \citet{dahn2008}, denoted $\alpha$, is related to the physical properties of the system as
\begin{equation}
\alpha = \rm{a}_{1}\left[1-\beta_{I}\left(M_{1}+M_{2}\right)/M_{2}\right]=0.276 AU
\end{equation}
Based on the parallax from \citet{dahn2008} the distance modulus for LSR1610 is $m-M=2.54\pm0.018$ and $\alpha$ is known in absolute units.
Combining the astrometric orbital period, eccentricity, and inclination with the RV semi-amplitude, $K_{1}$, directly determines the semi-major axis $a_{1}$. The spectroscopic mass function (eq. 5) and the relation for $\alpha$ are two equations in three unknowns ($M_{1}$, $M_{2}$, and $\beta_{I}$) since the spectroscopic mass function combined with the known inclination determines $M_{2}$ as a function of $M_{1}$. These relations together determine the I-band light ratio $\beta_{I}=\left(1+10^{0.4\Delta I}\right)^{-1}$ as a function of $M_{1}$. In Table 8 we list $M_2$ and $\beta_{I}$ derived from the combination of the astrometric and spectroscopic orbits for a range of $M_{1}$.

\begin{figure}
\plotone{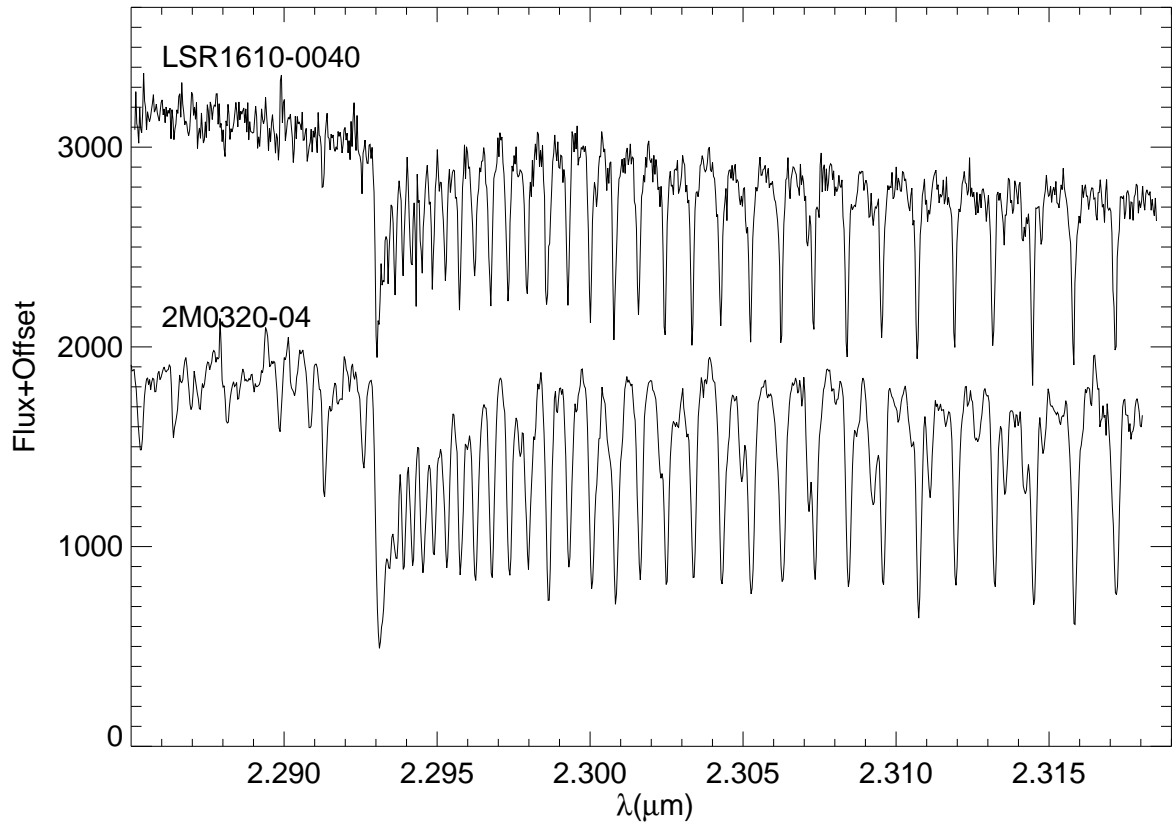}
\caption{NIRSPEC spectrum of LSR1610$-$0040 compared to the spectrum of the M8/L0 dwarf 2M0320$-$04. While the overall structure of the CO bandhead is similar, the lack of strong absorption features in LSR1610 blueward of the bandhead is possible evidence for low metallicity. }\label{lsr16spec}
\end{figure}

LSR1610 has been described as having  ``schizophrenic'' spectral properties and is in many ways an enigmatic object \citep{cushing2006, reiners2006}. The spectral features in the optical and NIR are indicative of a mildly metal poor mid-M dwarf with a number of unusual atomic absorption features. In Figure \ref{lsr16spec} we compare our high-resolution $K$-band spectra of this object to the spectrum of the M8/L0 dwarf 2M0320$-$04. While the overall
structure of the CO bandhead and the strengths of the  individual CO features is quite similar in these objects, we interpret the lack of absorption features blueward of the bandhead as possible evidence for low metallicity. \cite{dahn2008} found that the combined light of the system falls very near the locus of low-mass stars on color-magnitude diagrams (CMDs), both $M_{V}$ vs. $V-I$ and $M_{K_{s}}$ vs. $I-K_{s}$, meaning that it has roughly the expected luminosity and colors for an M dwarf. At the same time, they attribute a $B$ flux that is significantly suppressed to metal pollution by accretion of material from a hypothetical AGB star that could also be responsible for prominent Al lines in the NIR spectrum. Here we summarize what we know about this system and attempt to provide some possible explanations of the observed properties. 

\begin{itemize}

\item If we assume that the primary is indeed low in mass, say $0.075<M_1<0.2$M$_{\odot}$, then $0.138<\beta_{I}<0.177$. If we assume an average value of $\beta_{I}=\beta_{K_s}=0.16\pm0.02$ then the absolute magnitude of LSR1610A is $M_{K_s}=9.67\pm0.031$ mag. Our analysis constrains the mass ratio ($q=M_2/M_1$) of the system to be $0.83<q<1.4$ for $0.075<M_{1}<0.2$M$_{\odot}$.

\item MLRs are thought to be least sensitive to metallicity at NIR wavelengths. The updated $K$-band MLR from \citet{xia2009} would indicate a mass for LSR1610A of $M_1=0.103$M$_{\odot}$ after correcting for the light of LSR1610B, somewhat larger than the estimate based on the \citet{delfosse2000} relations of $M_1=0.088$M$_{\odot}$, though $M_{K}=9.67$ falls outside the range over which the \citet{delfosse2000} relation is formally defined.  

\item \citet{schilbach2009} found good general agreement between theoretical isochrones for old, low-metallicity stars from \citet{baraffe1997} and the absolute magnitudes of ten low-mass subdwarfs with measured parallaxes. Using slightly metal poor Lyon models ([M/H]=-0.5, t=10 Gyr) we estimated the mass of LSR1601A to be $M_{1}=0.095$M$_{\odot}$ and found a similar mass if the system is 1, 5 or 10 Gyr. These models also agree with the observed $J-K_{s}$ color of the combined light of this system. This object has a very large space velocity ($V_{tot}=243.3\pm2.0$ km s$^{-1}$), and based on the kinematic arguments in Section 7 we expect that this system is part of an older, thick disk population, making it unlikely that it is in fact very young. 

\end{itemize}

Based on the available observational evidence it remains difficult to explain LSR1610. From the empirical MLRs and theoretical isochrones it seems that 
LSR1610A is a low-mass star with $M_{1}\simeq0.1$M$_{\odot}$. For the mass range $M_{1}<0.1$M$_{\odot}$ the spectroscopic mass function requires that $M_1 \le M_2$, a somewhat implausible scenario if $\beta_{I}=0.16$ since the theoretical expectation is that the more massive component should be more luminous across the optical and infrared. The problem is alleviated somewhat if in fact LSR1610A is more massive than expected. Based on the available observational evidence we propose the following possible scenarios for LSR1610:

\begin{itemize}

\item The distance of LSR1610 could be underestimated. This would increase the mass estimate based on the isochrones and the MLRs and result in $M_{1}>M_{2}$. For example, if M$_{1}=0.2M_{\odot}$ then $M_{1}>M_{2}$ and the flux difference $\Delta I=1.99$~mag is consistent with theoretical models. This may seem an implausible explanation since \citet{dahn2008} and \citet{schilbach2009} find consistent parallaxes.

\item LSR1610B is a very unusual object. In principle LSR1610B could be an old compact object but would have to be very faint in $B$ ($M_{B}\sim22.4$) yet relatively bright in I band ($M_{I} \sim 14.3$, $B-I\sim8$). Based on the known properties of low-mass white dwarfs \citep{killic2007} we deem this scenario unlikely. \citet{dahn2008} reached a similar conclusion.

\end{itemize}

It is possible that in the near future this system could be resolved with aperture masking interferometry (e.g. \citealt{ireland2008}), resulting in a direct measurement of the masses of both components. We also carried out an empirical template analysis for this system, correcting the wavelength solutions for both the barycentric motion and the SB1 orbital solution. The expected velocity separation is rather large for this system ($\Delta RV\sim25$~km s$^{-1}$) and the flux ratio in K band is expected to be 5:1 making LSR1610 an excellent candidate for a double-lined system. Unfortunately, we have only one epoch of observations where the orbital phase is favorable for resolving the lines of the secondary and those spectra were obtained under poor observing conditions and are of relatively low S/N. We found no evidence for the lines of the secondary in the residuals of the fits at this epoch using an empirical template generated using the technique described in Section 5.1.

\subsection{2M1507$-$16}

This object is a very nearby mid-L dwarf with a parallax of $136.4\pm0.6$ mas \citep{dahn2002}. Our initial RV measurements of this object exhibited a statistically significant RV trend. We obtained additional NIRSPEC measurements under cloudy conditions in April and May, 2009, resulting in a total time baseline of more than six years. The amplitude of the RV variation is close to the limits of detection with our current technique, but we rule out constant RV [$P(\chi_{RV}^2\le)>0.999$]. The measurements, given in Table 9, are well-fit by a linear RV trend with a slope of $-72\pm23$~m s$^{-1}$ per year. Using a Monte-Carlo simulation we estimated the false alarm probability of observing a slope $|dRV/Dt|>72 $m s$^{-1}$ per year to be 2.2$\%$. This linear trend is also consistent with a single RV measurement from \citet{bailerjones2004} obtained in 2000, though this optical measurement has a very large error bar ($-39.3\pm1.5$ km s$^{-1}$ near HJD 2451661). \cite{basri2006} report a velocity
difference between 2000 and 2004 of 2.5 km s$^{-1}$ for this object, which is incompatible with the velocity trend seen 
in our data, though we note that these measurements also have large errors bars. We calculated the secular RV acceleration ($dRV/dt$) for this object following \citet{kurster2003} and estimated the maximum amplitude of this effect to be less than 0.6 m s$^{-1}$ yr$^{-1}$, far smaller than the observed trend.  The trend seen in our NIRSPEC observations could be indicative of a very long period system with $P>5000$ d, but we note that within our sub-sample of 46 objects observed on three or more epochs we may expect to observe at least one long-term slope with a false-alarm probability of $2.2\%$.

The mass of the L5 primary is poorly constrained based on photometry alone, but comparing absolute magnitudes and the temperature ($M_{K}=12.07$, $M_{bol}=15.32$, and $T_{\rm{eff}}=1703\pm60$ from \citealt{dahn2002}) to models from \citet{baraffe2003} and \citet{chabrier2000}, the minimum mass is expected to be in the range $0.06<M_{1}<0.07 M_{\sun}$ depending on the age of the system, assuming (based on the object's kinematics) $t>1 $Gyr. Based on the lack of Li absorption, \citet{reid2000} determined that this L dwarf has a minimum mass of $0.06M_{\sun}$. If we assume that the binary has near equal mass components, as is observed to be the case for many low-mass binaries \citep{allen2007}, with $M_{1}=M_{2}=0.065M_{\odot}$, then the mean orbital separation for a circular orbit would be $a>2.9$ AU if $P>5000$~d. At a distance of 7.3 pc this corresponds to an angular separation of $\theta>0.4\arcsec$. Unless the system is close to edge-on and observed at a very unfavorable phase, then such a binary would have been readily resolved in the high-resolution imaging observations presented by \citet{bouy2003} and \citet{reid2006} who found no luminous ($\Delta m<5$ mag) companions to 2M1507$-$16 over a range of separations $0.05\arcsec<\theta<1\arcsec$. Approximating the detection limit for faint companions from \citet{reid2006} as $\Delta M_{J}<5$ and assuming the 5 Gyr models from \citet{baraffe2003} for the secondary and 5 Gyr models from \citet{chabrier2000} for the primary then the non-detection constrains the mass ratio of the system to be $q<0.15$. It is also possible that the period of the system is much longer than 5000 d, resulting in an angular separation too large to be detected in the narrow fields of view of high-resolution imaging surveys. Using NIR imaging data obtained in 2005 and described in \citet{blake2008c} we detect only one source in an annulus of $5\arcsec$ to $20\arcsec$ down to limiting magnitude of $J\sim18.2$ and $K\sim16.3$. This source is also visible in archival 2MASS observations from 1998 and does not share the large ($0.9\arcsec$ year$^{-1}$; \citealt{dahn2002}) proper motion of 2M1507-16. 

\subsection{2M0746+20}

The orbital motion of the two components of this tight ($\theta\sim 150$ mas) binary system has been directly observed with high-resolution imaging, resulting in the measurement of the sum of the masses and the first dynamical mass estimate for an L dwarf \citep{bouy2004}. Based on high spatial resolution observations with HST, VLT, and Gemini, the orbital period of this system is estimated to be $P=3850.9^{+904}_{-767}$ days, so our NIRSPEC observations, shown in Figure \ref{2m0746}, span a considerable fraction of a period. The fit to the astrometric orbit determines the sum of the masses, $M_{1}+M_{2}=0.146^{+0.016}_{-0.006}M_{\sun}$, and the semi-major axis $a=2.53^{+0.37}_{-0.28}$ AU. The components of this system are resolved and the flux ratio of the components is measured to be 1.6:1 at $K$ band. Based on the orbital parameters determined in \citet{bouy2004} and the mass estimates from \citet{gizis2006}, the expected RV separation at periastron passage is $K_{1}+K_{_2}\simeq7$ km s$^{-1}$.  Using NIRSPEC with AO \citet{konopacky2010} resolved this system and obtained RV measurements for both components. Combining these measurements with the astrometric orbit yields direct measurements of the masses of both components, though for the case of 2M0746+20 the errors on the mass estimates are relatively large.  

While the flux ratio of the 
system is favorable for detecting the spectral lines of the secondary, the velocity separation is small compared to the NIRSPEC resolution, particularly for observations that do not coincide with the periastron passage in 2003. In this case where both components are of similar brightness, are expected to have similar spectral features, and are not resolved in velocity, we are effectively measuring the RV of the combined light of the system. If the components have exactly equal masses, spectral features, and luminosities, we would observe no RV variations at all. As a result of this combined light effect we do not necessarily expect to measure the true RV of 2M0746+20A using our technique. Using the orbital parameters from the astrometric orbit from \citet{bouy2004} we fit for $K_{1}$ and $\gamma$ assuming that the light from 2M0746+20B has a negligible impact on the measured RV and found $K_{1}=0.9\pm0.18$ km s$^{-1}$ and $\gamma=53.1\pm0.28$ km s$^{-1}$. The resulting RV orbit is shown in Figure \ref{2m0746}. We emphasize that these orbital parameters are very likely biased by the influence of 2M0746+20B, in particular our estimate of $K_{1}$ is expected to be too low. We carried out a simulation to assess the impact of a marginally-resolved secondary by fitting mock spectra that contained light from a secondary with a flux ratio of $F_{1}/F_{2}=1.5$ and velocity offsets of between $2<\Delta RV<10$ km s$^{-1}$ from the primary. We found that such companions significantly influenced the resulting RV measurement from fitting the combined light, $RV_{C}$, and that $|RV-RV_{C}|\approx \Delta RV\left(F_{2}/F_{1}\right)$, meaning that our RV measurements near periastron passage could be biased by up to a few km s$^{-1}$.

\section{Rotational and Space Velocities}

Our observations can also be used to study the statistics of the rotational and space velocities of UCDs. The rotational velocities of brown dwarfs and low-mass stars are important for understanding the angular momentum history of these objects. Measurements of a large number of UCD $V\sin{i}$ values exist in the literature, though the measurements were made using  techniques different from the one employed here. \citet{reiners2008} estimated rotational velocities by fitting the rotationally-broadened FeH stellar absorption features near $1~\mu$m using data from HIRES at Keck I and UVES at VLT. These optical data are of higher resolution than our NIRSPEC data resulting in $V\sin{i}$ sensitivity down to 3 km s$^{-1}$. Both \citet{bailerjones2004}, using VLT and the UVES spectrograph, and \citet{zapatero2007}, using Keck and NIRSPEC, follow the traditional technique of cross-correlating their spectra with a slowly rotating template and estimating $V\sin{i}$ from the width of the resulting cross correlation function. As shown in Figure \ref{vsin}, we find excellent overall agreement between our $V\sin{i}$ estimates and the values reported in the literature, including those in \citet{blake2007a}. 

 \citet{reiners2008} explored possible patterns in the rotation of M and L dwarfs as a function of spectral type in an effort to better understand the evolution of the angular momentum of UCDs as they age. The fact that both young and old L dwarfs are often rapid rotators indicates that their rotation differs from that of Sun-like stars. More massive stars lose angular momentum through solar winds, slowing as they age, but older field L dwarfs still appear to be relatively rapid rotators compared to Sun-like stars of similar ages. \citet{reiners2008} propose a braking mechanism that decreases in efficiency at the cooler atmospheric temperatures of UCDs and is very inefficient in mid- to late-L dwarfs, where the deceleration timescales become comparable to the age of the universe. At the cool temperatures of UCDs the stellar atmospheres are largely neutral, with few ions, reducing the strength of the coupling between the magnetic fields and the atmosphere and therefore reducing the strength of the braking mechanism. This model explains the lack of L dwarfs later than L3 found by \citet{reiners2008} to rotate slower than $V\sin{i}=20$~km s$^{-1}$. In Figure \ref{sptvvsin} we show our projected rotational velocities against spectral type as determined from optical diagnostics, the same spectral types used by \citet{reiners2008}, along with other measurements from the literature. We also see some evidence for increasing $V\sin{i}$ for objects with spectral types later than L3. We do, however, find four mid-L dwarfs that appear to be very slow rotators with $V\sin{i}<20$ km s$^{-1}$: 2M0355+11, 2M0500+03, 2M0835-08, and 2M1515+48. The spectra of these L dwarfs,  shown in Figure \ref{slow}, exhibit sharp, narrow CO features indicative of slow rotation and their spectra are relatively well fit by our theoretical UCD models. The L6 dwarf 2M1515+48 also has a very large space velocity, a potential indicator of old age. The \citet{reiners2008} sample includes approximately 15 L dwarfs with spectral types later than L3, comparable to our number. It is possible that these results are entirely consistent and that our slow rotators are simply observed close to pole-on. Comparing our $V\sin{i}$ measurements to those from \citet{reiners2008} for objects with spectral types later than or equal to L4 with a two-sided K-S test indicates a probability of $31\%$ that the two samples are drawn from the same underlying distribution.

Studying the kinematics of a population of stars, in particular their 3D motions, is a powerful tool for understanding the age of that population as well as for identifying subgroups that may have distinct histories from the main population. As a result of gravitational interactions, the velocity dispersion of a population of stars is expected to increase with time as it orbits in the galaxy. Models of UCDs, and brown dwarfs in particular, have an inherent mass-age degeneracy that makes it difficult to differentiate between a old low-mass star and young  brown dwarfs. This means that very few individual UCDs have reliable age estimates. Since UCDs are intrinsically very faint, the relatively bright UCDs we observe are necessarily close to the Sun and most have reliable proper motion, and sometimes parallax, measurements. Using these measurements, the statistical properties of the kinematics of the UCD population can be determined and the age of the population estimated using calibrated age-dispersion relations \citep{wielen1977}. \citet{faherty2009} present a analysis of the space velocities of 841 UCDs. In this work the authors do not have RV information but rely on photometric distance estimates combined with proper motions to estimate the tangential velocity, $V_{tan}$. As pointed out by \citet{reiners2009} the analysis by \citet{faherty2009} applies the results from \citet{wielen1977} by assuming $\sigma_{tot}=\sqrt{<V^{2}_{tot}>-<V_{tot}>^{2}}$ when in reality the dispersion of the total velocity must be determined from the dispersions of the individual velocity components.

We combined our radial velocities with proper motion and parallax measurements from the literature and photometric distance estimates from \citet{faherty2009} to estimate the total space velocity, $V_{tot}$, as well as $U$, $V$, $W$ components following the transformation presented by \citet{johnson1987}. We followed the right handed convention (positive $U$ is toward the galactic center) and we did not apply a correction to the local standard of rest. The distribution of $V_{tot}=\left(U^2+V^2+W^2\right)^{1/2}$ for 55 of our targets with distance and proper motion estimates from the literature is shown in Figure \ref{vtot} and the velocity components are listed in Table 10. The distribution of $V_{tot}$ is approximately Gaussian with a significant tail of targets with $V_{tot}>90$ km s$^{-1}$. Excluding these high velocity UCDs we found that the average space velocity of the UCDs in our sample is $\left<V_{tot}\right>=40.34\pm0.72$ km s$^{-1}$, similar to the mean L dwarf space velocity found by \citet{zapatero2007}. We estimated the velocity dispersions of our sample in each velocity component using a bootstrap simulation taking into account the reported errors on all of the measured quantities used to calculate $U$, $V$, and $W$. We calculated the $|W|$-weighted dispersions (Eqns. 1-3 in \citealt{wielen1977}) and estimated $\sigma_{tot}=\left(\sigma_{U}^2+\sigma_{V}^2+\sigma_{W}^2\right)^{1/2}$ excluding the high velocity UCDs. Based on the results of the bootstrap simulation we estimated $\sigma_{tot}=52.3\pm1.7$ km s$^{-1}$. We used Eqn. 16 from \citet{wielen1977} (Eqn. 3 from \citealt{reiners2009}) to estimate the age of our sample to 
be $5.0^{+0.7}_{-0.6}$ Gyr ($95\%$ confidence) not including any systematic uncertainty in the age-dispersion relation. This age is somewhat larger than the 3.1 Gyr estimated by \citet{reiners2009} for their sample of M7-M9.5 dwarfs, though they conclude that $10\%$ of their objects are actually young brown dwarfs. Our sample of mainly L dwarfs is lower in mass and may have less contamination from young objects, possbily contributing to the differences in the estimated ages of our two samples. Our results are in excellent agreement with those of \citet{seifahrt2010} who used a similar method to estimate the age for a sample of 43 L dwarfs. They also estimate the kinematic age of their sample to be approximately 5 Gyr and discuss the impliciations of this result. Since nearby M dwarfs appear to have an age of approximately 3 Gyr it is surprising that early L dwarfs, which, if below the hydrogen burning mass limit, will dim substanitally and move to later spectral types as they age, should be found to be older than M dwarfs. Possibly explanations for this proposed by \citet{seifahrt2010} include different initial velocity dispersions for M and L dwarfs, the influence of kinematic outliers in the estimation of the velocitiy dispersions, and non-Gaussianity in the velocity distributions.

The high velocity UCDs that we excluded from the $V_{tot}$ analysis are themselves potentially very interesting. Objects with high velocities may be subdwarfs from an older, lower-metallicity thick disk or halo population. In addition to LSR1610 we identified eight L dwarfs with $V_{tot}>90$ km s$^{-1}$ listed in Table 11. LSR1610 was one of the first known low-mass subdwarfs, and its $V_{tot}=243.3\pm2.0$ km s$^{-1}$ exceeds all of the rest of the objects in our sample and based on its velocity components it is considered to be a member of the halo population. In Figure \ref{thin} we compare the estimated UVW velocities to velocity ellipsoids for the thin disk and thick disk populations. While the majority of the UCDs in our sample have kinematics consistent with the local thin disk population, seven have large negative V velocities,  falling outside the $2\sigma$ velocity ellipsoid for the thin disk ($V<-55$ km s$^{-1}$), making them candidates for membership in the older thick disk population. 

\section{Tight Binary Rate}

Thanks mainly to direct imaging surveys, we now know that the population of UCD binaries has several significant differences from that of binaries composed of Sun-like stars. These differences are summarized in \citet{allen2007} and \citet{burgasser2007b}: $\bf{1:}$ UCD binaries tend to have mass ratios $q\sim1$ compared to $q\sim0.3$ for Sun-like stars, $\bf{2:}$ There are very few UCD binaries with $a>15-20$ AU, $\bf{3:}$ For UCDs the distribution of binary separation peaks around $a\sim7$ AU compared to $a\sim30$ AU for Sun-like stars, $\bf{4:}$ The overall fraction of binaries is $\sim20\%$ for UCDs compared to $60\%$ for Sun-like stars. Using a Bayesian approach that assumes a parameterized underlying distribution for the properties of UCD binary systems, \citet{allen2007} estimated that there is likely a small but significant population of small separation ($a<1$ AU) UCD binaries that are not detected by imaging surveys or the RV surveys that have been conducted to date. Systematic RV surveys like our NIRSPEC survey are important for constraining the binary fraction in this part of parameter space. An equal-mass UCD binary ($M_{1}=M_{2}=0.08$M$_{\sun}$) with an average separation of $a=1$ AU has an orbital period of 7 years and an RV semi-amplitude $K_{1}=K_{2}=1.6$ km s$^{-1}$, potentially detectable in our survey. We carried out a Monte Carlo simulation to estimate how efficiently our survey detects tight binaries and the overall rate of such systems. 

For our simulation we used a sub-sample of 40 targets from our survey meeting the following criteria: $K<12.5$, not a known binary, not a known subdwarf, observations on two or more epochs, and total time span of observations more then 350 days. Our simulation is similar in nature to the simulations of \citet{maxted2005}, though we made some simplifying assumptions. Using the actual times of observations of these 40 targets, as well as the actual RV error estimates for each observation, we generated $10^4$ hypothetical surveys where each of the 40 targets is a tight binary. For the mass ratio, $q=M_{2}/M_{1}$, we used a power-law distribution with exponent $\gamma=1.8$ \citep{allen2007}. We assumed that each of our UCDs has a mass of $0.08M_{\sun}$ and used a uniform distribution in $a$ between $0.01$ and $1.0$ AU to calculate spectroscopic binary orbits with random phase, argument of periapsis, $\omega$, and inclination. We assumed a distribution of eccentricity that is uniform between $e=0.0$ and $e=0.6$. We generated synthetic data based on the binary orbits, the actual times of the NIRSPEC observations, and adding noise based on the scaled error estimates at each epoch, $\sigma_{RV}$. Since we have already adjusted our empirical error estimates so that they are a good description of the overall statistical properties of our data, we feel that a synthetic data set generated in this way is a fair representation of a potential survey. We excluded epochs after April, 2009 since these observations were primarily for follow-up of suspected binary systems. To estimate the efficiency with which such systems would have been detected in our survey we counted the number of binary systems detected in each simulated survey with a conservative requirement that $P(\chi_{RV}^2\le)>0.999$ to qualify as a detection. Based on these simulations we estimated that our overall average efficiency in detecting systems with $0.01<a<1.0$ AU is $94\%$. We found that this efficiency is relatively insensitive to the assumed mass of the primary and, since it is peaked near 1.0, the assumed mass ratio distribution.

Our survey contains 43 targets that would have been included in the sample based on the above criteria alone and assuming that none of these systems passed the magnitude cut because they are unknown binaries. Within this sample we found one definite tight SB1 system (2M0320$-$04) and one possible SB1 system (2M1507$-$16), though if this system is a genuine SB1 then the separation may be larger than $a=1$ AU. The system 2M0746+20 was a known astrometric binary, with $a>1$AU, and LSR1610$-$0040 and LSR0602+39 were known subdwarfs. Based on the estimated detection efficiency, a flat prior on the binary fraction, and the detection of 1 (possibly 2) tight binaries, we use the binomial distribution to make a Bayesian estimate of the fraction of late-M and L dwarf systems in tight binaries ($90\%$ confidence) of $2.5^{+8.6}_{-1.6}\%$ ($5.0^{+9.6}_{-3.0}\%$). While our sample is small and the number of detected tight binaries is small, our observations are consistent with the overall binary properties from \citet{allen2007} and in particular with the prediction that $\sim3-4\%$ of UCDs are in binaries with $a<1$ AU. Assuming similar underlying statistical distributions to the ones we used here, \citet{maxted2005} combined radial velocity observations from the literature and found a relatively large number of close binaries, requiring an overall binary fraction of $40\%$, approximately a factor of two higher than those from \citet{allen2007}, \citet{joergens2008}, and \citet{basri2006}. While our results are consistent with the lower overall binary fraction, our current constraint does not rule out the higher value.

\subsection{Planet Detection Efficiency}

 For at least four reasons, UCDs are excellent targets for planet searches $\bf{1}$: The semi-amplitude
of the Doppler signal, $K_{1}$, increases inversely with the mass of the host
star ($K_{1} \propto M_{*}^{-2/3}$) for a fixed orbital period and
planetary mass. If the level of Doppler precision
achieved for F-G-K dwarfs can be achieved for UCDs, then it becomes
possible, in principle, to detect super-Earth
planets in the so called Habitable Zone (HZ), where in the absence of an atmosphere the equilibrium temperature of a planet is in the liquid water range. $\bf{2}$: UCDs have low luminosities compared
to Sun-like stars, meaning that the HZ moves closer in to the host,
serving to boost the Doppler signal of a planet of fixed mass beyond the
gain from the decrease in the mass of the primary. $\bf{3}$: Understanding the population of planetary companions to UCDs, and how this population differs from that of planets orbiting F-G-K dwarfs,  may allow astronomers to conduct important tests of theories of planet formation and migration. A significant population of giant planets orbiting  close to M and L dwarfs is not predicted by models of planet formation via core accretion (i.e. \citealt{laughlin2004, ida2005}), though such planets may be formed in the disk instability scenario \citep{boss2006}. $\bf{4}$: The ratio of the intrinsic
brightness of a planet to its host at thermal infrared wavelengths is orders of magnitude larger when
that host is a UCD rather than a Sun-like star. The relative proximity
to the Sun of the brightest UCDs that we observe also means that planetary
companions will have relatively large angular separations on the
sky. Favorable contrast ratios and angular separations mean that
planetary companions to UCDs will likely be among the best targets for
efforts to directly image and study the atmospheres of extrasolar
planets at infrared and mid-infrared wavelengths using high contrast imaging
techniques.

We ran a Monte Carlo simulation to estimate our ability to detect binary systems with small mass ratios ($q<0.2$) and orbital separations $0.01<a<1.0$ AU. While our current RV precision precludes the detection of sub-Jupiter-mass planets, we found that our survey is sensitive to giant planets and the lowest mass brown dwarfs over a wide range of orbital separations. For this simulation we assumed a uniform distribution in $a$ from 0.005 to 1.0 AU, circular orbits with $e=0$, random inclination, phase, and $\omega$, a primary mass of  $M_{1}=0.08$ M$_{\odot}$ and we tested a range of companion masses $0.5<M_{2}<15$ M$_{\rm{J}}$. We followed a similar procedure to that described previously in this section to generate $10^4$ realization of our survey and inserted spectroscopic orbits with random inclinations into the synthetic data and counted detections using the same $P(\chi_{RV}^2\le)>0.999$ criteria.  The results of this simulation are shown in Figure \ref{planet}.

With no definitive detections of companions in this mass regime we are only able to place upper limits on the prevalence of giant planetary companions to UCDs. We found that for small orbital separations, similar to those of the Hot Jupiters found orbiting Sun-like stars ($0.01<a<0.05$ AU), our survey is very sensitive to giant planets ($1<M_{2}<10$ M$_{\rm{J}}$) and the least massive brown dwarfs. Based on the average detection efficiencies, $P_{D}$, shown in Figure \ref{planet}, we estimate the following upper limits ($95\%$) for the rate of planetary companions, $f_{P}$, at orbital separations $a<0.05$~AU: for 1M$_{\rm{J}}$, $f_{p}<0.63$, for 2M$_{\rm{J}}$, $f_{p}<0.17$, for 5M$_{\rm{J}}$, $f_{p}<0.088$, for 10M$_{\rm{J}}$, $f_{p}<0.076$. It is also interesting to consider the case of companion with mass 15M$_{\rm{J}}$, possibly a Y dwarf, for which we find $f_{p}<0.089$ for $a<0.5$~AU. Based on other observational evidence we would expect the rate of close-in giant companions to be very low. If a system with $q<0.2$ and $a<0.05$~AU can be thought of as a binary system, as opposed to a planetary system, then the mass ratio distribution, separation distribution, and overall binary fraction from \citet{allen2007} would predict that the frequency of such companions would be exceedingly rare, orbiting fewer than 1 in $10^{6}$ UCDs. There are a number of very massive planets known to orbit at these distances from Sun-like stars, including WASP-19b \citep{hebb2010}, Corot-2b \citep{bouchy2008}, and WASP-14b \citep{joshi2008}, though these massive Hot Jupiters are relatively rare with a rate of occurrence of $f_{P}<0.005$ \citep{cummings2008,gould2006b}. The analyses of \citet{cummings2008} and \citet{johnson2007} both indicate that the overall rate of occurrence of planets decreases with stellar mass, with the rate of planets orbiting M stars being down by a factor of $\sim7$ compared to Sun-like stars. While there is no direct observational evidence pertaining to hot Jupiters orbiting UCDs, extrapolating from the observed properties of planets orbiting high mass stars indicates that the expected rate of occurrence, $f_{p}$, of these companions is probably less than $f_{P}<0.001$.

\section{Conclusions}

We report the results of a NIR Doppler survey designed to detect unseen companions to Ultracool Dwarfs. Infrared radial velocity measurements are a powerful tool for detecting UCD spectroscopic binaries and potentially even giant planets orbiting low-mass stars and brown dwarfs. We have developed a new technique for making radial velocity measurements at infrared wavelengths using CH$_{4}$
absorption features in the Earth's atmosphere as a simultaneous wavelength reference. We have used this technique to carry out a Doppler survey of UCDs using NIRSPEC on the Keck II telescope. We obtained more than 600 spectra of 59 UCDs on two or more epochs. We forward modeled our spectra using high-resolution theoretical UCD templates and the observed telluric spectrum, scaled for airmass, to measure radial velocity, RV, and the projected rotational velocity, $V\sin{i}$. We estimated our RV precision to be 50 m s$^{-1}$ for bright, slowly rotating M dwarfs and $100-300$ m s$^{-1}$ for slowly rotating UCDs. Within the framework of the photon-limited Doppler precision, we investigated different sources of noise that may limit our RV precision. Neglecting secular changes in CH$_{4}$ absorption features due to winds or changes is barometric pressure, we estimated that for our typical spectra (S/N$\sim75$), the telluric features themselves limit RV precision to $\sigma_{RV}>60$ m s$^{-1}$. We estimated that the overall precision of our RV measurements ($\sigma_{RV}$) is within a factor of two of the photon limit. There are several potential sources for this excess noise including changes in the symmetry of the NIRSPEC LSF, changes in the CH$_{4}$ absorption features themselves, episodes of high read noise in the detector, mismatch between the theoretical templates and the actual UCD spectra, and changes in the phase and amplitude of the ``fringing'' pattern seen in our NIRSPEC spectra. We explored avenues for mitigating these instrumental sources of RV variations, but ultimately were unable to improve the overall RV precision to the photon limit.

We identified one new spectroscopic binary, one candidate binary, and obtained RV measurements of two astrometric binaries from the literature. The candidate spectroscopic binary, 2M1507$-$16, shows a long-term RV trend and the absence of any observed companion in high-resolution imaging observations may indicate that the system is a low-amplitude spectroscopic binary with a period $P>2500$ d. Due to the proximity of the primary to the Sun, this object is an excellent target for ongoing efforts to directly image small mass ratio UCD binary systems. 

We also investigated the rotation and kinematics of the UCDs in our sample. Our estimates of $V\sin{i}$ are consistent with those in the literature. Unlike \cite{reiners2008}, our sample contains at least four mid-L dwarfs that are rotating at $V\sin{i}<15$ km s$^{-1}$, though we conclude that our two samples are statistically consistent. With proper motions and distance estimates from the literature, we were able to use our RV measurements to calculate the full 3D space velocities of 55 L dwarfs and estimate the width of the distribution of the measured values of $V_{tot}$. The dispersion of this velocity distribution is expected to be directly related to the age of a stellar population. We estimated the kinematic age of our sample to be $5.0^{+0.7}_{-0.6}$ Gyr. Based on simulations of our detection efficiency we estimated the rate of tight binaries ($a<1$ AU) to be $2.5_{-1.6}^{+8.6}\%$, consistent with an overall binary fraction of $20\%$ and a predicted tight binary fraction of $3-4\%$ given by \citet{allen2007}. Finally, we simulated our ability to detect giant planetary companions in short-period orbits and concluded that a large population of giant planets ($1<M_{2}<10$ M$_{\rm{J}}$) in close ($a<0.05$ AU) orbits would have been detected in our survey. Specifically, we find that the rate of ocurrence of companions with $M_{2}>1$M$_{\rm{J}}$ and $a<0.05$~AU is less than 62$\%$ and that the rate of ocurrence of more massive companions with $M_{2}>15$M$_{\rm{J}}$ and $a<0.5$~AU is less than 8.8$\%$

\textit{Acknowledgments}  We would like to thank D. Saumon and M. Marley for providing the library of high-resolution synthetic UCD spectra that
made this work possible as well as for thoughtful comments on this manuscript. We are grateful to the referee, A. Reiners, for his careful reading of our manuscript and his many helpful suggestions for improvements. We thank D. Spiegel
for several helpful discussions about statistics. It is a pleasure to acknowledge D. Latham, C. Stubbs, G. Torres for their guidance throughout this work. We also thank D. Finkbeiner, A. Loeb, and K. Stassun for helpful discussions and suggestions for improvements to this work. We also would like to thank J. Bailey for helpful discussions throughout the process of reducing and analyzing the NIRSPEC data. CHB acknowledges financial support from the Harvard Origins of Life Initiative, NExSci, and the NSF Astronomy \& Astrophysics Postdoctoral Fellowship program. Data presented herein were obtained at the W. M. Keck Observatory in part with telescope time allocated to the National Aeronautics and Space Administration through the agency’s scientific partnership with the California Institute of Technology and the University of California. The Observatory was made possible by the generous financial support of the W. M. Keck Foundation. We would like to thank G. Hill, J. Lyke and the Keck staff for their support over the course of this program. The Keck Observatory was made possible by the generous financial support of the W.M. Keck Foundation. The authors wish to recognize and acknowledge the very significant cultural role and reverence that the summit of Mauna Kea has always had within the indigenous Hawaiian community.  We are most fortunate to have the opportunity to conduct observations from this mountain. This research has benefited from the M, L, and T dwarf compendium housed at DwarfArchives.org and maintained by C. Gelino, D. Kirkpatrick, and A. Burgasser.

{\it Facilities:} \facility{Keck II}.

\clearpage

\clearpage

\begin{figure}
\plotone{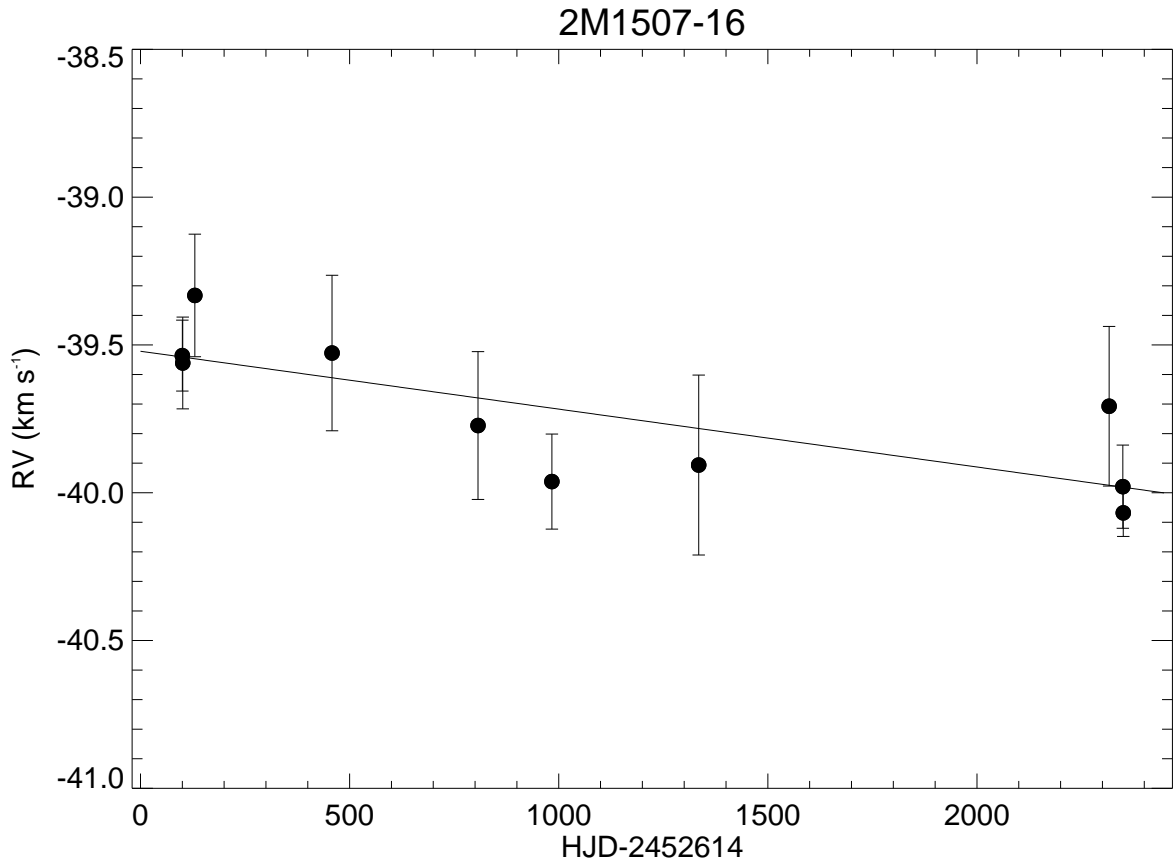}
\caption{RV measurements for 2M1507$-$16. The solid line is a linear RV trend that may be indicative of a very long period orbit. }\label{2m15}
\end{figure}

\begin{figure}
\plotone{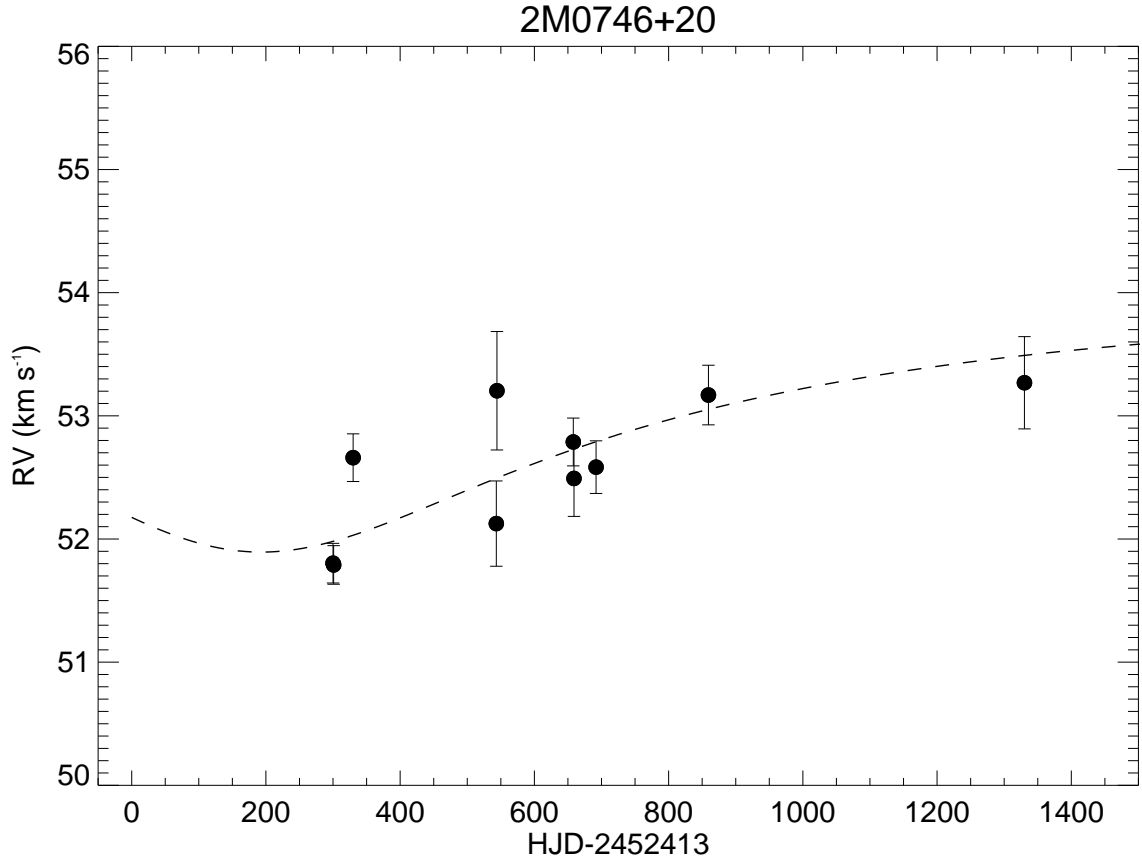}
\caption{RV measurements of 2M0746+20. The dashed line is the orbit of 2M0746+20A based on the astrometric observations of \citet{bouy2004} after fitting for $K_{1}$ and $\gamma$ with our RV measurements. The spectra of the two components of this system are not resolved in our NIRSPEC observations and so we measure the effective RV of the combined light.}\label{2m0746}
\end{figure}

\begin{figure}
\plotone{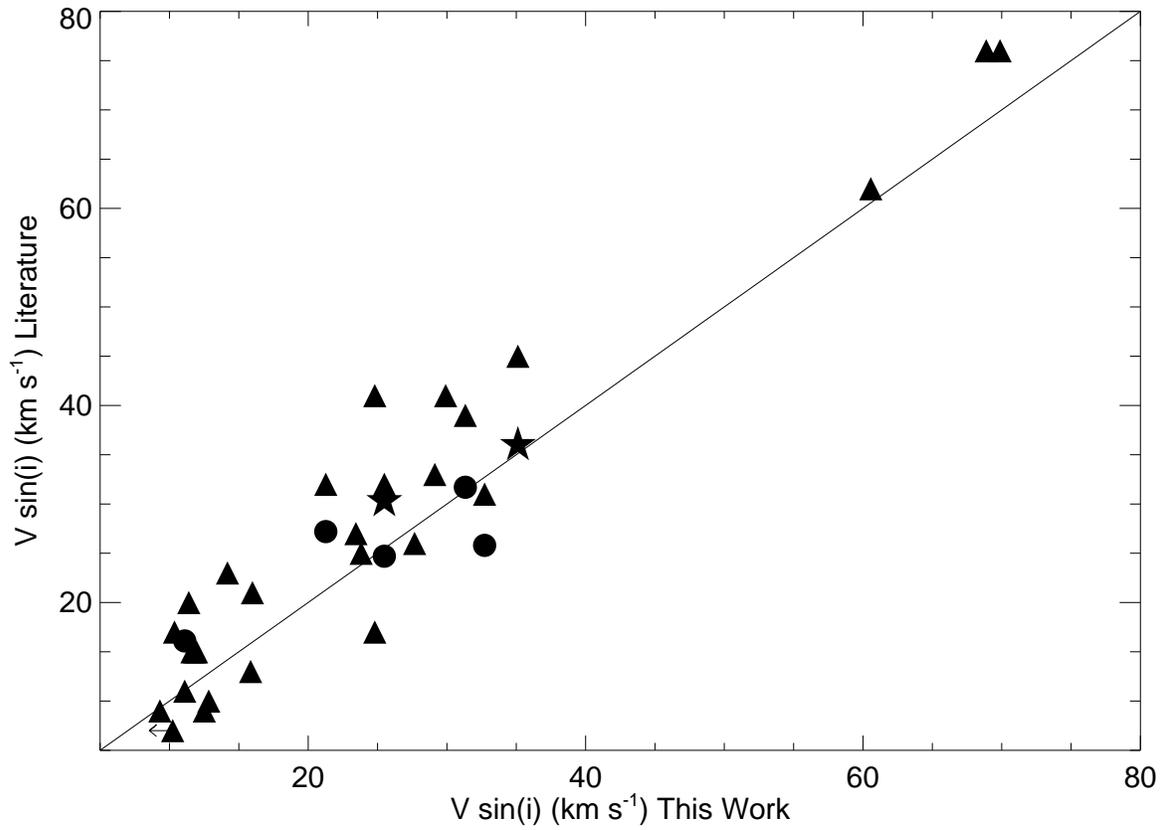}
\caption{Our $V\sin{i}$ measurements compared to those from the literature. Triangles are from \citet{reiners2008}, circles from \citet{bailerjones2004}, and stars from \citet{zapatero2007}. While there is significant scatter in the $V\sin{i}$ estimates we find overall good agreement between our measurements and those from the literature.} \label{vsin}
\end{figure}

\begin{figure}
\plotone{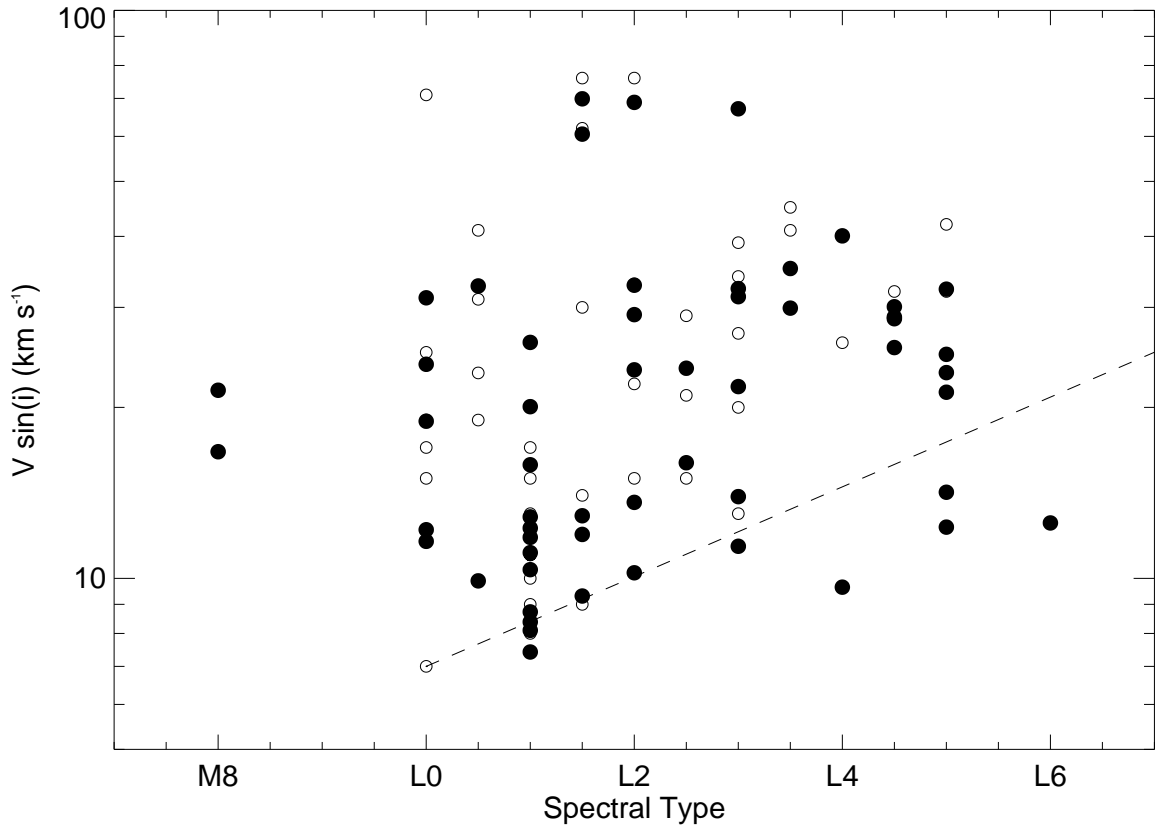}
\caption{Optical spectral type compared to our measured $V\sin{i}$ (solid points) along with those from \citet{reiners2008} (open circles). The dashed line corresponds to the $V\sin{i}$ lower envelope suggested by \citet{reiners2008}. We found four examples slowly-rotating ($V\sin{i}<15$~km s$^{-1}$) objects with mid-L spectral types.}\label{sptvvsin}
\end{figure}

\begin{figure}
\plotone{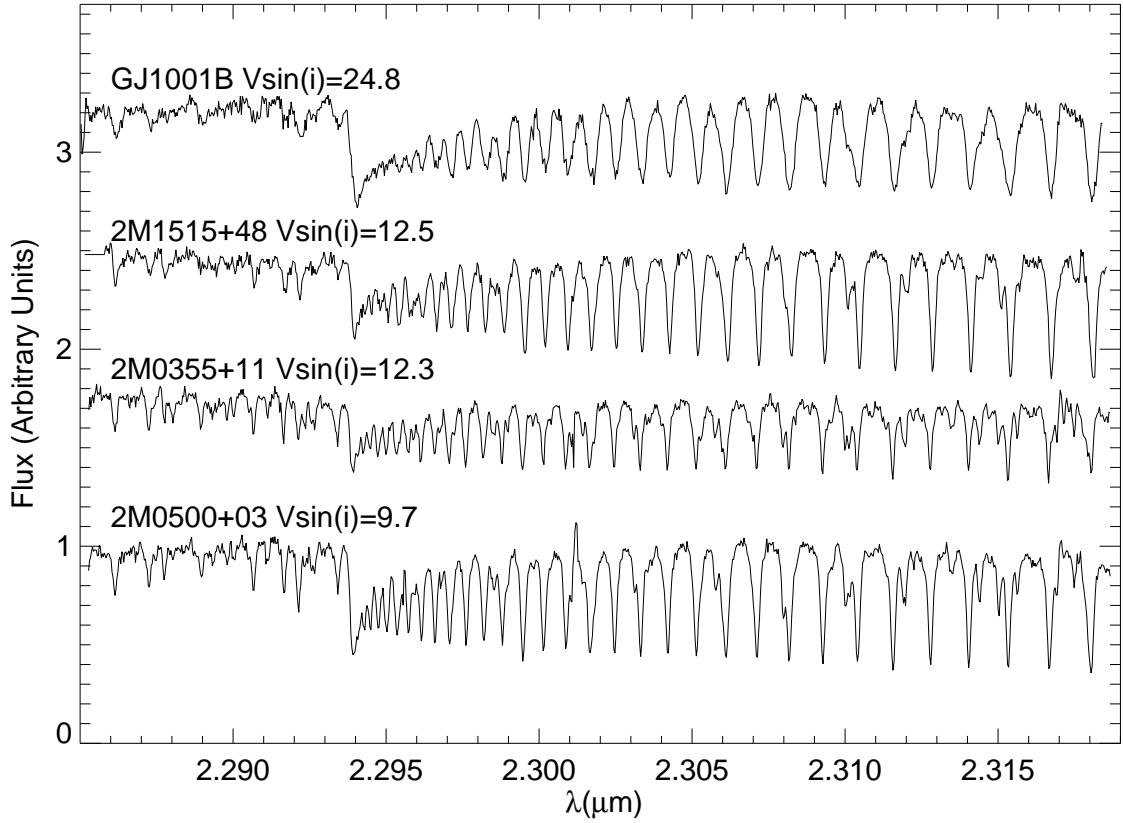}
\caption{NIRSPEC spectra of three slowly rotating mid-L dwarfs after removal of the telluric features. For comparison, the L5 dwarf GJ 1001B with $V\sin{i}=24.8\pm0.4$ km s$^{-1}$ is also shown. 2M1515+48 (L6), 2M0355+11 (L5), and 2M0500+03(L4) all appear to have projected rotation velocities smaller than that of GJ 1001B.}\label{slow}
\end{figure}

\begin{figure}
\plotone{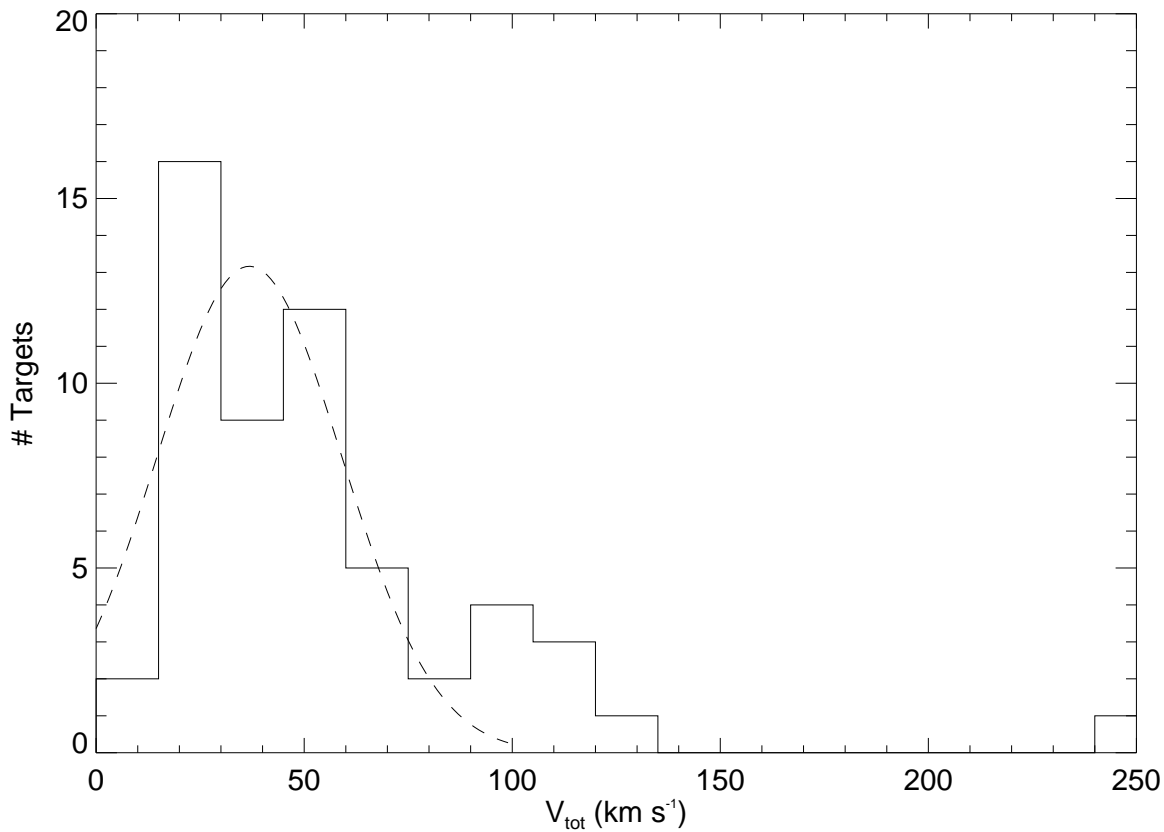}
\caption{Distribution of 3D space velocities ($V_{tot}$) of the objects in our sample along with a best-fit Gaussian distribution. Excluding the population of high velocity L dwarfs with $V_{tot}>90$~km s$^{-1}$, we estimate the width of the $|W|$-weighted $V_{tot}$ distribution to be $\sigma_{V_{tot}}=52.3\pm1.7$~km s$^{-1}$. The high velocity UCD LSR1610$-$0040 is shown near 250 km s$^{-1}$. We note that $\sigma_{tot}=\left(\sigma_{U}^2+\sigma_{V}^2+\sigma_{W}^2\right)^{1/2} \ne \sqrt{<V^{2}_{tot}>-<V_{tot}>^{2}}$. }\label{vtot}
\end{figure}

\begin{figure}
\plotone{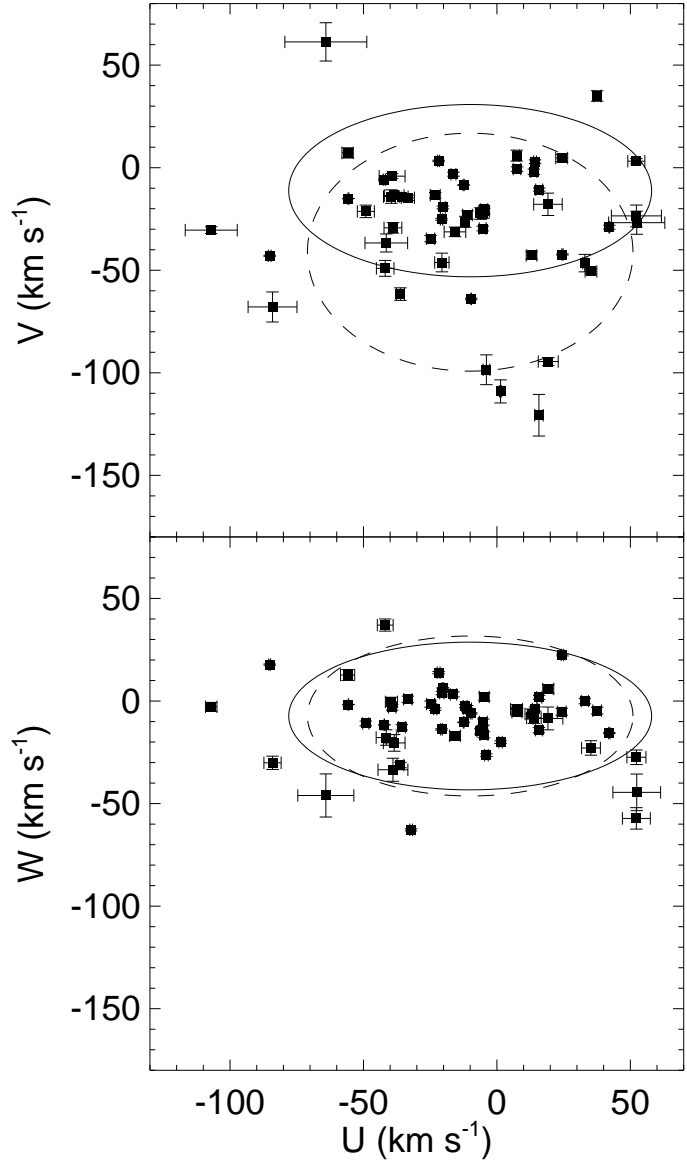}
\caption{Velocities of our targets in the U-V and U-W planes. The solid line is the $2\sigma$ velocity ellipsoid for the thin disk population and the dashed line is the $1\sigma$ ellipsoid for the thick disk population \citep{binney1998}. Large negative $V$ velocities are a possible indication of thick disk membership.}\label{thin}
\end{figure}

\begin{figure}
\plotone{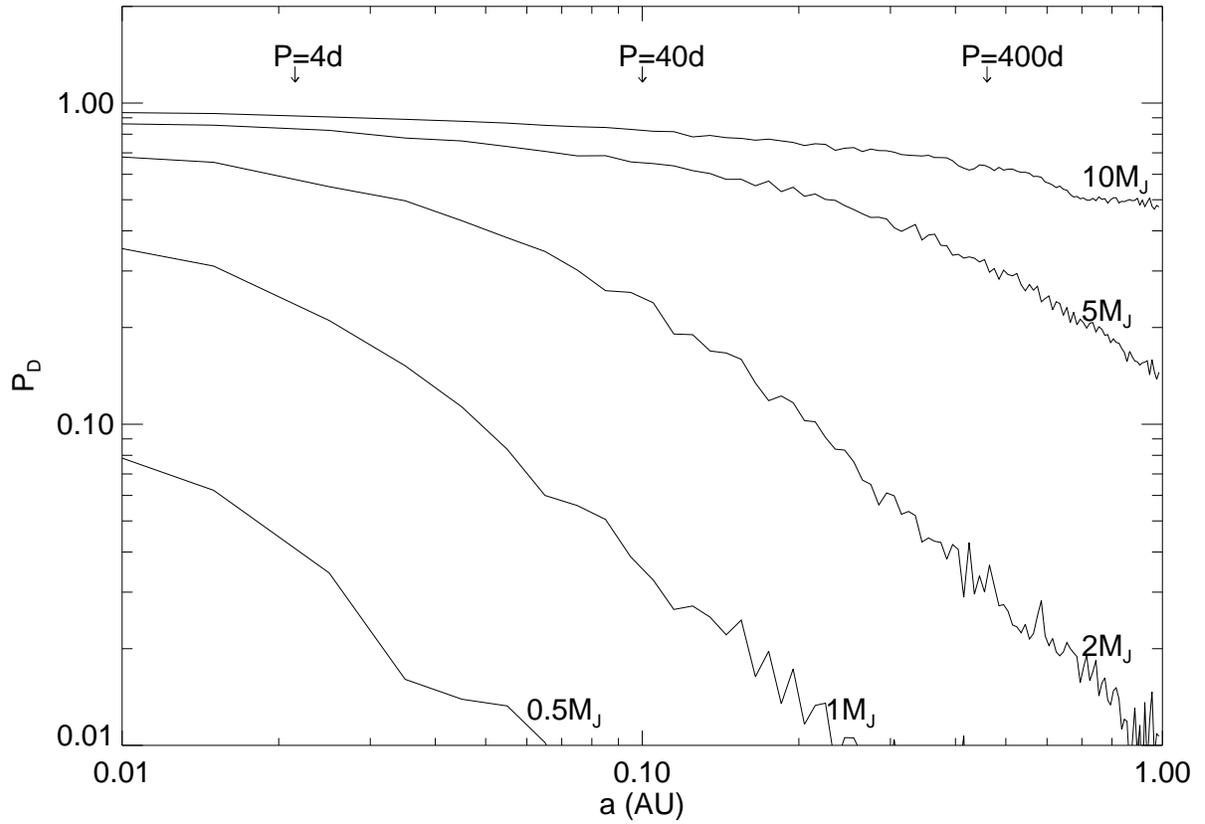}
\caption{Results of a Monte Carlo simulation designed to estimate the planet detection efficiency of our survey, $P_{D}$. We have sensitivity to Jupiter-mass companions at small orbital separations and the most massive planets out to $a\sim1.0$~AU. The Roche limit for these systems is $a>0.002$ AU \citep{blake2008c}. }\label{planet}
\end{figure}

\begin{deluxetable}{lccccc}
\tabletypesize{\scriptsize}
\tablecaption{L Dwarf Sample}
\tablewidth{0pt}
\tablehead{
\colhead{ID} & \colhead{RA} & \colhead{DEC} & \colhead{J} & \colhead{K} &\colhead{Sp.Type}
}
\startdata
  GJ1001B  &   00 04 34.84   &    $-$40 44 05.8  &  13.11  &  11.40  &    L5\\
    2M0015+35  &   00 15 44.76   &    +35 16 02.6  &  13.88  &  12.26  &    L2\\
    2M0036+18  &   00 36 16.17   &    +18 21 10.4  &  12.47  &  11.06  &  L3.5\\
    2M0045+16  &   00 45 21.43   &    +16 34 44.6  &  13.06  &  11.37  &    L2\\
    2M0141+18  &   01 41 03.21   &    +18 04 50.2  &  13.88  &  12.49  &    L1\\
    2M0144$-$07  &   01 44 35.36   &    $-$07 16 14.2  &  14.19  &  12.27  &    L5\\
    2M0213+44  &   02 13 28.80   &    +44 44 45.3  &  13.49  &  12.21  &  L1.5\\
    2M0227$-$16  &   02 27 10.36   &    $-$16 24 47.9  &  13.57  &  12.14  &    L1\\
    2M0228+25  &   02 28 11.01   &    +25 37 38.0  &  13.84  &  12.47  &   L0\\
    2M0235$-$23  &   02 35 59.93   &    $-$23 31 20.5  &  12.69  &  12.19  &    L1\\
    2M0251$-$03  &   02 51 14.90   &    $-$03 52 45.9  &  13.06  &  11.66  &    L3\\
    2M0306$-$36  &   03 06 11.59   &    $-$36 47 52.8  &  11.69  &  10.63  &    M8\\
    2M0320$-$04  &   03 20 28.39   &    $-$04 46 35.8  &  13.26  &  12.13  &    M8/L0\\
    2M0355+11  &   03 55 23.37   &    +11 33 43.7  &  14.05  &  11.53  &    L5\\
    2M0500+03  &   05 00 21.00   &    +03 30 50.1  &  13.67  &  12.06  &    L4\\
    2M0523$-$14  &   05 23 38.22   &    $-$14 03 02.2  &  13.08  &  11.64  &  L2.5\\
    2M0539$-$00  &   05 39 52.00   &    $-$00 59 01.9  &  14.03  &  12.53  &    L5\\
    2M0543+64  &   05 43 18.87   &    +64 22 52.8  &  13.57  &  12.05  &    L1\\
   LSR0602+39  &   06 02 30.45   &    +39 10 59.2  &  12.30  &  10.86  &    L1\\
    2M0632+83  &   06 32 06.17   &    +83 05 01.4  &  13.75  &  12.40  &  L0.5\\
    2M0652+47  &   06 52 30.73   &    +47 10 34.8  &  13.51  &  11.69  &  L4.5\\
    2M0700+31  &   07 00 36.64   &    +31 57 26.6  &  12.92  &  11.32  &  L3.5\\
    2M0717+57  &   07 17 16.26   &    +57 05 43.0  &  14.64  &  12.94  &    L3\\
    2M0746+20  &   07 46 42.56   &    +20 00 32.1  &  11.76  &  10.47  &  L0.5\\
    2M0828$-$13  &   08 28 34.19   &    $-$13 09 19.8  &  12.80  &  11.30  &    L2\\
    2M0835$-$08  &   08 35 42.56   &    $-$08 19 23.7  &  13.17  &  11.14  &    L5\\
    2M0847$-$15  &   08 47 28.72   &    $-$15 32 37.2  &  13.51  &  12.06  &    L2\\
    2M0921$-$21  &   09 21 14.10   &    $-$21 04 44.6  &  12.78  &  11.69  &  L1.5\\
    2M0911+74    &   09 11 12.97  &      +74 01 08.1   & 12.92   &  11.75  &  L0\\
    2M1022+58  &   10 22 48.21   &    +58 25 45.3  &  13.50  &  12.16  &    L1\\
    2M1045$-$01  &   10 45 24.00   &    $-$01 49 57.6  &  13.16  &  11.78  &    L1\\
    2M1048+01  &   10 48 42.81   &    +01 11 58.0  &  12.92  &  11.62  &    L1\\
    2M1108+68  &   11 08 30.81   &    +68 30 16.9  &  13.12  &  11.58  &    L1\\
    2M1112+35  &   11 12 25.67   &    +35 48 13.1  &  14.58  &  12.72  &  L4.5\\
    2M1155$-$37  &   11 55 39.52   &    $-$37 27 35.0  &  12.81  &  11.46  &    L2\\
    2M1203+00  &   12 03 58.12   &    +00 15 50.0  &  14.01  &  12.48  &    L3\\
    2M1221+02  &   12 21 27.70   &    +02 57 19.8  &  13.17  &  11.95  &    L0\\
    2M1300+19  &   13 00 42.55   &    +19 12 35.4  &  12.72  &  11.62  &    L1\\
    2M1305$-$25  &   13 05 40.19   &    $-$25 41 05.9  &  13.41  &  11.75  &    L2\\
    2M1425$-$36  &   14 25 27.98   &    $-$36 50 22.9  &  13.75  &  11.81  &   L3\\
    2M1439+19  &   14 39 28.36   &    +19 29 14.9  &  12.76  &  11.55  &    L1\\
    2M1506+13  &   15 06 54.41   &    +13 21 06.0  &  13.36  &  11.74  &    L3\\
    2M1507$-$16  &   15 07 47.69   &    $-$16 27 38.6  &  12.83  &  11.31  &    L5\\
    2M1515+48  &   15 15 00.83   &    +48 47 41.6  &  14.11  &  12.50  &    L6\\
    2M1539$-$05  &   15 39 41.89   &    $-$05 20 42.8  &  13.92  &  12.57  &   L4\\
    2M1552+29  &   15 52 59.06   &    +29 48 48.5  &  13.48  &  12.02  &    L0\\
    2M1555$-$09  &   15 55 15.73   &    $-$09 56 05.5  &  12.56  &  11.44  &    L1\\
 LSR1610$-$0040  &   16 10 29.00   &    $-$00 40 53.0  &  12.91  &  12.02  &  sdM7\\
    2M1645$-$13  &   16 45 22.11   &    $-$13 19 51.6  &  12.45  &  11.15  &  L1.5\\
    2M1658+70  &   16 58 03.80   &    +70 27 01.5  &  13.29  &  11.91  &    L1\\
    2M1705$-$05  &   17 05 48.34   &    $-$05 16 46.2  &  13.31  &  12.03  &  --\\
    2M1731+27  &   17 31 29.74   &    +27 21 23.3  &  12.09  &  10.91  &    L0\\
    2M1807+50  &   18 07 15.93   &    +50 15 31.6  &  12.93  &  11.60  &  L1.5\\
    2M1821+14  &   18 21 28.15   &    +14 14 01.0  &  13.43  &  11.65  &  L4.5\\
    2M1854+84  &   18 54 45.97   &    +84 29 47.1  &   11.54  &   11.35  &    -- \\
    2M2036+10  &   20 36 03.16   &    +10 51 29.5  &  13.95  &  12.45  &    L3\\
    2M2057$-$02  &   20 57 54.09   &    $-$02 52 30.2  &  13.12  &  11.72  &  L1.5\\
    2M2104$-$10  &   21 04 14.91   &    $-$10 37 36.9  &  13.84  &  12.37  &  L2.5\\
    2M2224$-$01  &   22 24 43.81   &    $-$01 58 52.1  &  14.07  &  12.02  &  L4.5\\
     
\enddata
\tablecomments{Coordinates, magnitudes, and spectral types of the targets in our sample as gathered from www.dwarfarchives.com and cross checked against the database of UCDs recently published by \citet{faherty2009}. The spectral types listed here are those derived from optical diagnostics. Optical spectral types are not available for 2M1705$-$05 and 2M1854+84 but based on their broad-band colors these objects are expected to be L dwarfs.}

\end{deluxetable}

\begin{deluxetable}{lccc}
\tabletypesize{\scriptsize}
\tablecaption{Table of RV Measurements\label{tbl-1}}
\tablewidth{0pt}
\tablehead{
\colhead{Object} & \colhead{HJD-2400000} & \colhead{RV} & \colhead{$\sigma$}\\
\colhead{} & \colhead{} & \colhead{km s$^{-1}$} & \colhead{km s$^{-1}$}
}
\startdata

     GJ1001B   &  53271.94   &    32.95   &  0.70\\
      GJ1001B   &  53272.88   &    33.28   &  0.38\\
      GJ1001B   &  53328.74   &    32.83   &  0.43\\
      GJ1001B   &  53669.82   &    32.60   &  0.27\\
      ..........  & & &\\

\enddata
\tablecomments{The remainder to Table 2 is available on-line.}
\end{deluxetable}

\begin{deluxetable}{lccccccc}
\tabletypesize{\scriptsize}
\tablecaption{Measured Properties of L dwarfs}
\tablewidth{0pt}
\tablehead{
\colhead{ID} & \colhead{N Obs.} & \colhead{$\Delta$ T} & \colhead{T$_{\rm{eff}}$} & \colhead{V $\sin{i}$} & \colhead{$<RV>$} & \colhead{RMS} & \colhead{P($\chi_{RV}^{2}\le$)}\\
\colhead{} & \colhead{} & \colhead{days} & \colhead{K} & \colhead{km s$^{-1}$} & \colhead{km s$^{-1}$} & \colhead{km s$^{-1}$} & \colhead{}
}
\startdata
 
   GJ1001B  &   4  & 398 &  1600  &    24.80$\pm$ 0.40  &    32.84$\pm$ 0.17  &   0.30  &     0.47  \\
    2M0015+35  &   5 & 406  &  2200  &    10.23$\pm$ 2.55  &   $-$37.35$\pm$ 0.16  &   0.37  &     0.17  \\
    2M0036+18  &   6  & 351 &   2000  &    35.12$\pm$ 0.57  &    19.02$\pm$ 0.15  &   0.26  &     0.05  \\
    2M0045+16  &   6  &  351 & 2100  &    32.82$\pm$ 0.17  &     3.29$\pm$ 0.17  &   0.40  &     0.05  \\
    2M0141+18  &   4  &  822 & 2200  &             $<$9.0  &    24.65$\pm$ 0.13  &   0.39  &     0.76  \\
    2M0144$-$07  &   5  &  821 & 1600  &    23.03$\pm$ 0.45  &    $-$2.55$\pm$ 0.10  &   0.36  &     0.97  \\
    2M0213+44  &   5  &  823 & 2200  &    12.89$\pm$ 2.34  &   $-$23.47$\pm$ 0.11  &   0.20  &     0.52  \\
    2M0227$-$16  &   2  & 431 &  2100  &             $<$9.0  &    48.58$\pm$ 0.15  &   0.25  &     0.49  \\
    2M0228+25  &   5  &  406 & 2100  &    31.19$\pm$ 0.81  &    23.07$\pm$ 0.21  &   0.29  &     0.13  \\
    2M0235$-$23  &   6  &  823 & 2200  &    15.85$\pm$ 0.53  &    15.38$\pm$ 0.11  &   0.26  &     0.71  \\
    2M0251$-$03  &   3  & 351 & 2100  &    21.76$\pm$ 0.38  &     1.26$\pm$ 0.13  &   0.31  &     0.80  \\
    2M0306$-$36  &   2  &  358 & 1900  &    21.44$\pm$ 0.26  &    11.44$\pm$ 0.19  &   0.18  &     0.43  \\
    2M0320$-$04  &  16  &  1180 & 1900  &    16.71$\pm$ 0.53  &     3.13$\pm$ 0.06  &   5.32  &     $>$0.99  \\
    2M0355+11  &   2  &  322 & 1900  &    12.31$\pm$ 0.15  &    11.92$\pm$ 0.22  &   0.45  &     0.50  \\
    2M0500+03  &   2  &  322 & 1600  &     9.65$\pm$ 0.36  &    15.94$\pm$ 0.16  &   0.09  &     0.26  \\
    2M0523$-$14  &   5  &  821 & 2100  &    15.98$\pm$ 0.31  &    12.21$\pm$ 0.09  &   0.19  &     0.70  \\
    2M0539$-$00  &   4  &  614 & 1600  &    32.30$\pm$ 0.75  &    13.91$\pm$ 0.15  &   0.38  &     0.83  \\
    2M0543+64  &   2  &  57 & 2200  &    20.06$\pm$ 0.52  &    18.64$\pm$ 0.23  &   0.11  &     0.26  \\
   LSR0602+39  &  10  &  1029 & 2100  &    12.51$\pm$ 0.22  &     7.94$\pm$ 0.05  &   0.24  &     $>$0.99  \\
    2M0632+83  &   2  &  96 & 1900  &     9.90$\pm$ 0.40  &   $-$26.00$\pm$ 0.24  &   0.06  &     0.11  \\
    2M0652+47  &  10  &  1030 & 1600  &    30.08$\pm$ 1.76  &     $-$7.03$\pm$ 0.07  &   0.21  &     0.33  \\
    2M0700+31  &   6  &  670 & 2100  &    29.91$\pm$ 0.27  &   $-$42.42$\pm$ 0.09  &   0.25  &     0.77  \\

    2M0717+57  &   2  &  1 & 2100  &    13.93$\pm$ 0.55  &   $-$16.32$\pm$ 0.17  &   0.19  &     0.54  \\
    2M0746+20  &  10  &  1030 & 2100  &    32.72$\pm$ 0.56  &    52.37$\pm$ 0.06  &   0.59  &     $>$0.99  \\
    2M0828$-$13  &   5  &  357 & 2300  &    29.13$\pm$ 5.00  &    25.85$\pm$ 0.08  &   0.28  &     0.93  \\
    2M0835$-$08  &   8  &  1028 & 2200  &    14.18$\pm$ 0.43  &     29.89$\pm$ 0.06  &   0.15  &     0.52  \\

    2M0847$-$15  &   5  &  999 & 2200  &    23.29$\pm$ 0.32  &     2.02$\pm$ 0.10  &   0.08  &     0.03  \\
    2M0911+74  &   2  &  362 & 2000  &    12.18$\pm$ 0.59  &    $-$4.06$\pm$ 0.15  &   0.28  &     0.79  \\
    2M0921$-$21  &   3  &  680 & 2000  &    11.95$\pm$ 0.49  &    80.53$\pm$ 0.11  &   0.05  &     0.04  \\

    2M1022+58  &   3  & 682 &  2200  &    11.81$\pm$ 0.21  &    19.29$\pm$ 0.11  &   0.10  &     0.08  \\
    2M1045$-$01  &   4  &  677 & 2200  &             $<$9.0  &     6.31$\pm$ 0.10  &   0.22  &     0.77  \\
    2M1048+01  &   7  &  1030 & 2100  &    10.36$\pm$ 0.25  &    24.25$\pm$ 0.06  &   0.22  &     0.91  \\

    2M1108+68  &   5  & 999 & 2100  &    26.03$\pm$ 0.29  &    $-$9.84$\pm$ 0.11  &   0.29  &     0.49  \\
    2M1112+35  &   6  & 2249 & 2100  &    28.65$\pm$ 1.01  &    $-$4.28$\pm$ 0.13  &   0.39  &     0.33  \\
    2M1155$-$37  &   2  & 264 &  2200  &    13.61$\pm$ 0.31  &    45.57$\pm$ 0.11  &   0.27  &     0.73  \\
    2M1203+00  &   3  &  678 & 1600  &    31.33$\pm$ 0.55  &    $-$0.22$\pm$ 0.16  &   0.57  &     0.88  \\

    2M1221+02  &   3  & 681 &  2000  &    23.82$\pm$ 0.42  &    $-$8.79$\pm$ 0.14  &   0.32  &     0.49  \\
    2M1300+19  &   5  &  1040 & 2200  &    12.83$\pm$ 2.10  &   $-$17.60$\pm$ 0.12  &   0.22  &     0.28  \\
    2M1305$-$25  &   2  &  29 &2100  &    68.88$\pm$ 2.60  &     6.37$\pm$ 0.35  &   0.05  &     0.06  \\
    2M1425$-$36  &   2  &  303 & 2000  &    32.37$\pm$ 0.66  &     5.37$\pm$ 0.25  &   0.25  &     0.38  \\
    2M1439+19  &   5  &  1069 & 2200  &    11.10$\pm$ 0.24  &   $-$26.74$\pm$ 0.09  &   0.17  &     0.43  \\
    2M1506+13  &   3  &  853 & 2200  &    11.39$\pm$ 0.94  &    $-$0.68$\pm$ 0.11  &   0.12  &     0.28  \\
    2M1507$-$16  &   9  &  2249 & 1600  &    21.27$\pm$ 1.86  &   $-$39.85$\pm$ 0.05  &   0.27  &     $>$0.99  \\

    2M1515+48  &   4  &  707 & 1600  &    12.52$\pm$ 1.65  &   $-$29.97$\pm$ 0.11  &   0.14  &     0.13  \\

    2M1539$-$05  &   3  &  29 & 1800  &    40.09$\pm$ 0.76  &    27.33$\pm$ 0.24  &   0.48  &     0.52  \\
    2M1552+29  &   4  &  1237 & 2000  &    18.91$\pm$ 0.57  &   $-$18.43$\pm$ 0.11  &   0.03  &     0.01  \\
    2M1555$-$09  &   5  &  1233 & 2200  &             $<$9.0  &    14.84$\pm$ 0.10  &   0.25  &     0.46  \\
 LSR1610$-$0040\tablenotemark{a}  &   4  & 1541 &  1900  &    16.84$\pm$ 3.11  &   $-$97.89$\pm$ 0.21  &   5.41  &     $>$0.99  \\
    2M1645$-$13  &   6  & 1234 &  2200  &     9.31$\pm$ 0.27  &    26.58$\pm$ 0.06  &   0.15  &     0.51  \\
    2M1658+70  &   5  &  2041 & 2200  &    12.26$\pm$ 0.76  &   $-$25.60$\pm$ 0.12  &   0.17  &     0.20  \\
    2M1705$-$05  &   3  &  425 & 2200  &    27.67$\pm$ 0.32  &    12.19$\pm$ 0.11  &   0.11  &     0.18  \\
    2M1731+27  &   3  &  526 & 2000  &    11.62$\pm$ 0.15  &   $-$29.76$\pm$ 0.11  &   0.34  &     0.92  \\
    2M1807+50  &   4  &  528 & 2100  &    69.88$\pm$ 2.48  &    $-$0.36$\pm$ 0.46  &   1.14  &     0.45  \\
    2M1821+14  &   2  &  162 & 2200  &    28.85$\pm$ 0.16  &     9.78$\pm$ 0.16  &   0.14  &     0.43  \\
    2M1854+84  &   3  &  351 & 2000  &    10.23$\pm$ 0.13  &    $-$2.93$\pm$ 0.17  &   0.10  &     0.10  \\
    2M2036+10  &   2  &  146 & 2100  &    67.11$\pm$ 1.52  &    19.66$\pm$ 0.47  &   1.59  &     0.86  \\
    2M2057$-$02  &   3  &  351 & 2100  &    60.56$\pm$ 0.37  &   $-$24.68$\pm$ 0.43  &   0.61  &     0.48  \\
    2M2104$-$10  &   6  &  748 & 2200  &    23.44$\pm$ 0.23  &   $-$21.09$\pm$ 0.12  &   0.39  &     0.58  \\
    2M2224$-$01  &   8  &  750 & 1600  &    25.49$\pm$ 0.41  &   $-$37.55$\pm$ 0.09  &   0.19  &     0.09  \\

\enddata

\tablecomments{Measured properties of the L dwarfs in our sample. $T_{\rm{eff}}$ is based on the best-fit synthetic template with $\log{g}=5.0$ (cgs). The errors on $V\sin{i}$ are estimated from the scatter in fits to multiple observations. The errors on the systemic velocity $<RV>$ are also determined from the scatter in the fits to multiple observations. The value of $P(\chi_{RV}^{2}\le)$ is the probability of getting a smaller value of $\chi_{RV}^{2}$ and assumes the null hypothesis of no intrinsic RV variations.}
\tablenotetext{a}{For this object we used a template with $\log{g}=5.5$.}
\end{deluxetable}

\begin{deluxetable}{lcc}
\tabletypesize{\scriptsize}
\tablecaption{RV Measurements of 2M0320$-$04\label{tbl-1}}
\tablewidth{0pt}
\tablehead{
\colhead{HJD-2400000} & \colhead{RV} & \colhead{$\sigma$}\\
\colhead{} & \colhead{km s$^{-1}$} & \colhead{km s$^{-1}$}
}
\startdata
52921.102 &  6.148 & 0.378\\
52922.113  & 6.550 & 0.294\\
52957.027 &  4.392 & 0.327\\
53272.125 & $-$6.443 & 0.234\\
53273.086 & $-$6.461 & 0.235\\
53328.828 & $-$4.113 & 0.225\\
53421.719 &  6.274 & 0.194\\
53669.887 &  5.905 & 0.291\\
53670.879 &  6.363 & 0.191\\
53686.859 &  5.453 & 0.178\\
53742.809  &$-$2.912 & 0.331\\
53743.856 & $-$3.140 & 0.258\\
53744.840 & $-$3.378 & 0.407\\
54023.969 & $-$7.276 & 0.216\\
54100.746 &  1.134 & 0.295\\
54101.754 &  1.398 & 0.673\\

\enddata
\end{deluxetable}

\begin{deluxetable}{lc}
\tabletypesize{\scriptsize}
\tablecaption{Orbital and System Parameters for 2M0320$-$04}
\tablewidth{0pt}
\tablehead{
\colhead{Parameter} & \colhead{Value} 
}
\startdata
 
Period (day) & 246.9 $\pm0.52$ \\
e  &    0.067$\pm0.015$\\
$\omega$ ($^{\circ}$) & 167.2$\pm18.7$\\
$T_{0}$ (MJD) & 53529.9$\pm11.8$\\
$K_{1}$ (km s$^{-1}$) & 6.92$\pm$0.12\\
$\gamma$ (km s$^{-1}$) & 0.154$\pm0.072$\\
$M_{2}\sin{i}$ ($M_{\sun}$) & 0.2032$\left(M_{1}+M_{2}\right)^{2/3}\pm0.0007$\\
$a_{1}\sin{i}$ (AU) & 0.157$\pm$0.003\\

\enddata
\tablecomments{Estimated orbital and system parameters for 2M0320$-$04. These results are consistent with the earlier estimates from \citet{blake2008b}}. 
\end{deluxetable}

\begin{deluxetable}{lcc}
\tabletypesize{\scriptsize}
\tablecaption{RV Measurements of LSR1610$-$0040\label{tbl-1}}
\tablewidth{0pt}
\tablehead{
\colhead{HJD-2400000} & \colhead{RV} & \colhead{$\sigma$}\\
\colhead{} & \colhead{km s$^{-1}$} & \colhead{km s$^{-1}$}
}
\startdata

53421.176  & -95.041 & 0.436\\
53431.455\tablenotemark{a}   & -95 & 1 \\ 
53597.766  & -97.629 & 0.266\\
53800.835\tablenotemark{b}   & -108.1 & 1.6\\
53948.773  & -95.828 & 0.726\\
54962.938  & -106.482 & 0.686\\
\enddata

\tablenotetext{a}{\citet{basri2006}}
\tablenotetext{b}{\citet{dahn2008}}

\end{deluxetable}

\begin{deluxetable}{lc}
\tabletypesize{\scriptsize}
\tablecaption{Orbital and System Parameters for LSR16010$-$0040\label{tbl-1}}
\tablewidth{0pt}
\tablehead{
\colhead{Parameter} & \colhead{Value} 
}
\startdata
 
Period (day; fixed) & 607.1$\pm4.34$ \\
e (fixed) &    0.444$\pm0.017$\\
$i$ ($^{\circ}$; fixed) & 83.2$\pm1.0$\\
$\omega$ ($^{\circ}$; fixed) & 151.4$\pm4.6$\\
$T_{0}$ (MJD) & 53680.2  $\pm8.2$\\
$K_{1}$ (km s$^{-1}$) & 8.02$\pm0.90$\\
$\gamma$ (km s$^{-1}$) & $-$99.49$\pm0.41$\\
$M_{2}^3$ ($M_{\sun}$) & 0.025$\left(M_{1}+M_{2}\right)^{2}\pm0.008$\\
$a_{1}$ (AU) & 0.404$\pm0.042$

\enddata
\tablecomments{Estimated orbital and system parameters for LSR1610$-$0040 assuming the astrometric orbital parameters from \citet{dahn2008}. Using four NIRSPEC measurements and one other from the literature we fit for the time of periastron passage, $T_{0}$, the systemic velocity, $\gamma$, and the RV semi-amplitude, $K_{1}$.}
\end{deluxetable}

\begin{deluxetable}{ccc}
\tabletypesize{\scriptsize}
\tablecaption{LSR1610}
\tablewidth{0pt}
\tablehead{
\colhead{$M_{1}$} & \colhead{$M_{2}$} & \colhead{$\beta_{I}$}\\
\colhead{$M_\odot$} & \colhead{$M_\odot$} & \colhead{}
}
\startdata
0.20 & 0.144 & 0.138\\
0.15 & 0.124 & 0.150\\
0.10 & 0.10  & 0.166\\
0.095 & 0.098 & 0.168\\
0.090 & 0.096 & 0.170\\
0.085 & 0.093 & 0.172\\
0.08  & 0.900 & 0.175\\
0.075 & 0.087 & 0.177\\
0.07  & 0.085 & 0.181\\

\enddata
\tablecomments{System parameters from the combined analysis of the astrometric and spectroscopic orbits for a range of $M_{1}$. The mass of the primary uniquely determines the mass of the secondary, $M_{2}$, and the I-band flux ratio, $\beta_{I}=l_{2}/(l_{1}+1{2})$.}
\end{deluxetable}

\begin{deluxetable}{lcc}
\tabletypesize{\scriptsize}
\tablecaption{RV Measurements of 2M1507$-$16\label{tbl-1}}
\tablewidth{0pt}
\tablehead{
\colhead{HJD-2400000} & \colhead{RV} & \colhead{$\sigma$}\\
\colhead{} & \colhead{km s$^{-1}$} & \colhead{km s$^{-1}$}
}
\startdata
 52714.066  &$-$39.536  &0.120\\
 52715.105  &$-$39.561  &0.155\\
 52744.043  &$-$39.333  &0.207\\
 53072.102  &$-$39.527  &0.263\\
 53421.086  &$-$39.773  &0.250\\
 53597.730  &$-$39.962  &0.161\\
 53948.742  &$-$39.906  &0.304\\
 54930.055  &$-$39.707  &0.270\\
 54962.895  &$-$39.979  &0.141\\
 54963.895  &$-$40.068  &0.080\\

\enddata
\end{deluxetable}

\begin{deluxetable}{lcccc}
\tabletypesize{\scriptsize}
\tablecaption{UVW Velocity Components}
\tablewidth{0pt}
\tablehead{
\colhead{ID} & \colhead{U} & \colhead{V} & \colhead{W} & \colhead{V$_{\rm{tot}}$}\\
\colhead{} & \colhead{km s$^{-1}$} & \colhead{km s$^{-1}$} & \colhead{km s$^{-1}$} & \colhead{km s$^{-1}$}
}
\startdata
 
    GJ1001B  &      1.37$\pm$ 0.42  &  $-$109.16$\pm$ 5.55  &   $-$20.03$\pm$ 0.62  &   110.99$\pm$ 5.61  \\
      2M0015+35  &     13.03$\pm$ 2.02  &   $-$42.88$\pm$ 1.76  &    $-$6.53$\pm$ 2.54  &    45.29$\pm$ 3.69  \\
      2M0036+18  &    $-$42.15$\pm$ 0.41  &    $-$5.80$\pm$ 0.22  &   $-$11.59$\pm$ 0.09  &    44.10$\pm$ 0.47  \\
      2M0045+16  &    $-$22.96$\pm$ 1.79  &   $-$13.41$\pm$ 1.30  &    $-$3.98$\pm$ 0.59  &    26.88$\pm$ 2.29  \\
      2M0141+18  &    $-$48.92$\pm$ 3.08  &   $-$21.11$\pm$ 2.97  &   $-$10.94$\pm$ 1.08  &    54.39$\pm$ 4.41  \\
      2M0144$-$07  &    $-$20.82$\pm$ 2.62  &   $-$46.41$\pm$ 4.43  &     3.88$\pm$ 0.84  &    51.01$\pm$ 5.22  \\
      2M0213+44  &     19.54$\pm$ 5.37  &   $-$17.52$\pm$ 5.40  &    $-$8.81$\pm$ 5.69  &    27.68$\pm$ 9.51  \\
      2M0227$-$16  &    $-$36.24$\pm$ 1.55  &   $-$61.75$\pm$ 3.01  &   $-$31.17$\pm$ 1.21  &    78.09$\pm$ 3.60  \\
      2M0228+25  &    $-$39.56$\pm$ 2.87  &   $-$14.08$\pm$ 3.28  &    $-$0.38$\pm$ 2.15  &    41.99$\pm$ 4.86  \\
      2M0235$-$23  &    $-$12.45$\pm$ 0.20  &    $-$8.45$\pm$ 0.16  &   $-$10.38$\pm$ 0.16  &    18.28$\pm$ 0.30  \\
      2M0251$-$03  &     15.68$\pm$ 1.58  &  $-$120.09$\pm$10.09  &   $-$14.14$\pm$ 1.26  &   121.93$\pm$10.29  \\
      2M0320$-$04  &     52.03$\pm$10.22  &   $-$26.46$\pm$ 5.39  &   $-$44.23$\pm$ 8.68  &    73.24$\pm$14.45  \\
      2M0352+02  &    $-$55.97$\pm$ 2.22  &     7.41$\pm$ 2.53  &    12.77$\pm$ 2.73  &    57.89$\pm$ 4.34  \\
      2M0355+11  &     $-$6.29$\pm$ 1.06  &   $-$22.19$\pm$ 2.91  &   $-$14.79$\pm$ 1.32  &    27.40$\pm$ 3.37  \\
      2M0500+03  &     $-$4.82$\pm$ 1.34  &   $-$20.41$\pm$ 1.79  &   $-$16.43$\pm$ 1.48  &    26.64$\pm$ 2.68  \\
      2M0523$-$14  &    $-$16.27$\pm$ 0.93  &    $-$3.26$\pm$ 0.95  &     3.30$\pm$ 1.17  &    16.92$\pm$ 1.77  \\
      2M0539$-$00  &    $-$21.86$\pm$ 0.37  &     3.31$\pm$ 0.32  &    13.84$\pm$ 0.55  &    26.09$\pm$ 0.74  \\
     LSR0602+39  &    $-$11.90$\pm$ 0.38  &   $-$26.51$\pm$ 2.59  &    $-$2.47$\pm$ 0.74  &    29.16$\pm$ 2.72  \\
      2M0652+47  &      7.38$\pm$ 0.93  &     5.97$\pm$ 2.65  &    $-$4.08$\pm$ 2.16  &    10.34$\pm$ 3.55  \\
      2M0700+31  &     42.13$\pm$ 0.15  &   $-$28.97$\pm$ 0.95  &   $-$15.62$\pm$ 0.25  &    53.46$\pm$ 1.00  \\
      2M0717+57  &     13.88$\pm$ 1.17  &    $-$2.03$\pm$ 1.34  &    $-$8.07$\pm$ 2.77  &    16.19$\pm$ 3.29  \\
      2M0746+20  &    $-$55.63$\pm$ 0.14  &   $-$15.04$\pm$ 0.07  &    $-$1.74$\pm$ 0.09  &    57.65$\pm$ 0.19  \\
      2M0828$-$13  &    $-$38.71$\pm$ 3.50  &   $-$13.49$\pm$ 1.10  &   $-$20.68$\pm$ 3.91  &    45.91$\pm$ 5.36  \\
      2M0835$-$08  &    $-$33.22$\pm$ 2.48  &   $-$14.77$\pm$ 1.33  &     0.89$\pm$ 1.65  &    36.37$\pm$ 3.27  \\
      2M0847$-$15  &     15.84$\pm$ 1.28  &   $-$10.68$\pm$ 0.74  &     1.79$\pm$ 1.01  &    19.19$\pm$ 1.79  \\
      2M0921$-$21  &     19.43$\pm$ 3.85  &   $-$94.59$\pm$ 2.04  &     5.77$\pm$ 2.12  &    96.74$\pm$ 4.84  \\
      2M0953$-$10  &     $-$4.72$\pm$ 1.83  &   $-$21.15$\pm$ 1.47  &     2.14$\pm$ 1.80  &    21.77$\pm$ 2.96  \\
      2M1022+58  &    $-$65.37$\pm$14.65  &    60.85$\pm$ 9.29  &   $-$46.79$\pm$10.47  &   100.82$\pm$20.26  \\
      2M1045$-$01  &    $-$35.71$\pm$ 2.41  &   $-$14.22$\pm$ 1.00  &   $-$12.83$\pm$ 1.33  &    40.52$\pm$ 2.93  \\
      2M1048+01  &    $-$24.91$\pm$ 1.57  &   $-$34.50$\pm$ 1.47  &    $-$1.43$\pm$ 1.47  &    42.58$\pm$ 2.61  \\
      2M1051+56  &    $-$15.81$\pm$ 4.00  &   $-$31.47$\pm$ 2.39  &   $-$16.95$\pm$ 1.99  &    39.09$\pm$ 5.06  \\
      2M1108+68  &    $-$41.59$\pm$ 8.42  &   $-$36.59$\pm$ 4.47  &   $-$17.93$\pm$ 4.02  &    58.23$\pm$10.34  \\
      2M1112+35  &    $-$20.75$\pm$ 0.45  &   $-$25.17$\pm$ 0.50  &   $-$13.83$\pm$ 0.31  &    35.43$\pm$ 0.74  \\
      2M1155$-$37  &     35.42$\pm$ 2.09  &   $-$50.55$\pm$ 1.19  &   $-$22.97$\pm$ 3.41  &    65.86$\pm$ 4.17  \\
      2M1203+00  &    $-$83.82$\pm$ 8.69  &   $-$67.49$\pm$ 7.03  &   $-$29.92$\pm$ 3.11  &   111.70$\pm$11.60  \\
      2M1221+02  &     51.86$\pm$ 9.23  &   $-$23.29$\pm$ 5.19  &   $-$27.25$\pm$ 3.49  &    63.04$\pm$11.15  \\
      2M1300+19  &     $-$3.87$\pm$ 1.22  &   $-$98.51$\pm$ 7.00  &   $-$26.40$\pm$ 0.67  &   102.06$\pm$ 7.14  \\
      2M1305$-$25  &    $-$20.27$\pm$ 0.86  &   $-$19.39$\pm$ 0.67  &     6.24$\pm$ 0.33  &    28.74$\pm$ 1.14  \\
      2M1425$-$36  &     $-$5.05$\pm$ 1.09  &   $-$22.47$\pm$ 2.23  &   $-$10.26$\pm$ 1.50  &    25.22$\pm$ 2.90  \\
      2M1439+19  &    $-$85.11$\pm$ 0.53  &   $-$42.92$\pm$ 0.28  &    17.75$\pm$ 0.31  &    96.96$\pm$ 0.68  \\
      2M1506+13  &    $-$41.64$\pm$ 3.22  &   $-$48.84$\pm$ 3.82  &    36.85$\pm$ 2.92  &    74.01$\pm$ 5.79  \\
      2M1507-16   & $-$26.41$\pm$0.14 &  $-$17.10$\pm$0.12 &  $-$39.67$\pm$0.12  &  50.63$\pm$0.22  \\      
     2M1515+48  &   $-$107.23$\pm$ 9.50  &   $-$30.47$\pm$ 1.63  &    $-$2.76$\pm$ 2.11  &   111.51$\pm$ 9.86  \\
      2M1539$-$05  &     37.58$\pm$ 1.20  &    35.00$\pm$ 2.43  &    $-$4.76$\pm$ 1.62  &    51.58$\pm$ 3.16  \\
      2M1552+29  &    $-$11.09$\pm$ 1.80  &   $-$22.92$\pm$ 1.96  &    $-$3.75$\pm$ 1.54  &    25.74$\pm$ 3.08  \\
      2M1555$-$09  &     52.11$\pm$ 3.08  &     3.19$\pm$ 1.25  &   $-$57.00$\pm$ 5.03  &    77.29$\pm$ 6.03  \\
   LSR1610$-$0040  &     $-$32.11$\pm$ 0.82  &  $-$232.9$\pm$ 1.7  &  $-$62.75$\pm$ 0.51  &   243.3$\pm$2.0  \\
      2M1645$-$13  &     33.00$\pm$ 0.81  &   $-$46.83$\pm$ 4.46  &    $-$0.08$\pm$ 1.36  &    57.28$\pm$ 4.74  \\
      2M1658+70  &     24.56$\pm$ 0.23  &   $-$42.29$\pm$ 0.29  &    22.43$\pm$ 0.45  &    53.81$\pm$ 0.59  \\
      2M1705$-$05  &     14.25$\pm$ 0.53  &     2.95$\pm$ 0.88  &    $-$3.80$\pm$ 0.98  &    15.05$\pm$ 1.42  \\
      2M1731+27  &     $-$5.28$\pm$ 1.35  &   $-$29.71$\pm$ 1.25  &   $-$13.28$\pm$ 1.14  &    32.97$\pm$ 2.16  \\
      2M1807+50  &      7.46$\pm$ 1.41  &    $-$0.67$\pm$ 1.12  &    $-$5.41$\pm$ 1.88  &     9.24$\pm$ 2.61  \\
      2M2036+10  &     24.26$\pm$ 2.30  &     4.73$\pm$ 1.75  &    $-$5.27$\pm$ 1.68  &    25.27$\pm$ 3.34  \\
      2M2104$-$10  &    $-$38.79$\pm$ 3.43  &   $-$29.37$\pm$ 2.65  &   $-$33.36$\pm$ 5.82  &    58.99$\pm$ 7.26  \\
      2M2224$-$01  &     $-$9.62$\pm$ 0.11  &   $-$63.97$\pm$ 0.39  &    $-$6.01$\pm$ 0.34  &    64.97$\pm$ 0.52  \\

\enddata

\tablecomments{UVW velocity components and $V_{tot}$ for the portion of our sample with distance estimates and proper motions in \citet{faherty2009}. We use a right handed system with $U>0$ toward the galactic center. Errors are derived from a Monte Carlo simulation that includes the errors on radial velocity, distance, and proper motion.}
\end{deluxetable}

\begin{deluxetable}{lccc}
\tabletypesize{\scriptsize}
\tablecaption{High Velocity UCDs\label{tbl$-$1}}
\tablewidth{0pt}
\tablehead{
\colhead{Name} & \colhead{$V_{tot}$} & \colhead{$V$} & \colhead{V$\sin{i}$}\\
\colhead{} & \colhead{km s$^{-1}$}  & \colhead{km s$^{-1}$} & \colhead{km s$^{-1}$}
}
\startdata

GJ1001B    &  110.99$\pm$5.61 &   $-$109.16$\pm$5.55 & 24.80$\pm$ 0.40   \\
2M0227$-$16   &78.09$\pm3.60$ & $-$61.75$\pm3.01$ & $<9$\\
2M0251$-$03   &   121.93$\pm$10.29 &  $-$120.09$\pm10.09$ &  21.76$\pm$ 0.38   \\
2M0921$-$21  &    96.74$\pm$4.84  & $-$94.59$\pm2.04$ & 11.95$\pm$ 0.49  \\
2M1022+58   &   100.82$\pm$20.26  & 60.85$\pm9.29$ & 11.81$\pm$ 0.21    \\
2M1203+00   &   111.70$\pm$11.60  & $-$67.49$\pm7.03$ & 31.33$\pm$ 0.55      \\
2M1300+19   &   102.06$\pm$7.14   & $-$98.51$\pm7.00$ & 12.83$\pm$ 2.10       \\
2M1439+19   &   96.96$\pm$0.68    & $-$42.92$\pm0.28$ & 11.10$\pm$ 0.24       \\
2M1515+48   &   111.1$\pm$9.86   & $-$30.47$\pm1.63$ & 12.52$\pm$ 1.65        \\
LSR1610$-$0040  &    243.30$\pm$2.00 &  $-$232.96$\pm1.70$ & 16.84$\pm$ 3.11\\
2M2224$-$01   & 64.97$\pm0.52$ & $-$63.97$\pm0.39$ & 25.49$\pm0.41$\\

\enddata

\tablecomments{High velocity ($V_{tot}>90$~km s$^{-1}$ or $V<-55$ km s$^{-1}$) UCDs identified in our NIRSPEC survey. These objects have space velocities exceeding the velocities of the local stellar population, indicating that they may be part of a distinct, older thick disk population. One of the high velocity targets, LSR1610$-$0040, is an L subdwarf.}
\end{deluxetable}

\end{document}